\documentclass[prd,nofootinbib,preprint,showpacs,showkeys,superscriptaddress]{revtex4}
\usepackage[english]{babel}
\usepackage{amsmath,amssymb}
\usepackage[dvips]{graphicx}
\usepackage{ifthen,array}
\usepackage[sort&compress]{natbib}
\usepackage{amsfonts}
\usepackage{bbm}

\DeclareMathAlphabet{\mathsc}{OT1}{cmr}{m}{sc}

\allowdisplaybreaks

\newcommand{\nn}  {\nonumber}
\def\vb#1{\vbox to #1 pt{}}

\def\10{$SO(10)$}
\def\21{SU(2) $\otimes$ U(1) }

\def\422{$SU(4) \otimes SU(2) \otimes SU(2)$}
\def\321{SU(3) $\otimes$ SU(2) $\otimes$ U(1)}

\def\lsim{\raise0.3ex\hbox{$\;<$\kern-0.75em\raise-1.1ex\hbox{$\sim\;$}}}
\def\gsim{\raise0.3ex\hbox{$\;>$\kern-0.75em\raise-1.1ex\hbox{$\sim\;$}}}
\def\vev#1{\left\langle #1\right\rangle}

\newcommand{\flux}[2][]{\ensuremath{\ifthenelse{\equal{#1}{}}{}{^{#1}\!}\mathit{#2}}}

\newcommand{\AddrAHEP}{%
  AHEP Group, Instituto de F\'{\i}sica Corpuscular --
  C.S.I.C./Universitat de Val{\`e}ncia \\
  Edificio Institutos de Paterna, Apt 22085, E--46071 Valencia, Spain}
\newcommand{\AddrLisb}{%
 Departamento de F\'\i sica and CFIF, Instituto Superior T\'ecnico\\
          Av. Rovisco Pais 1, $\:\:$ 1049-001 Lisboa, Portugal }
\begin{document}

\preprint{IFIC/04-40}

\vspace*{2cm} \title{Invisible Higgs Boson Decays in Spontaneously
  Broken R-Parity }

\author{M.~Hirsch} \email{mahirsch@ific.uv.es}\affiliation{\AddrAHEP}
\author{J.~C.~Romao}
\email{jorge.romao@ist.utl.pt}\affiliation{\AddrLisb}
\author{J.~W.~F.~Valle} \email{valle@ific.uv.es}
\affiliation{\AddrAHEP} \author{A.~Villanova del Moral}
\email{Albert.Villanova@ific.uv.es}\affiliation{\AddrAHEP}


\begin{abstract}
  
  The Higgs boson may decay mainly to an invisible mode characterized
  by missing energy, instead of the Standard Model channels. This is a
  generic feature of many models where neutrino masses arise from the
  spontaneous breaking of ungauged lepton number at relatively low
  scales, such as spontaneously broken R-parity models.  Taking these
  models as framework, we reanalyze this striking suggestion in view
  of the recent data on neutrino oscillations that indicate non-zero
  neutrino masses. We show that, despite the smallness of neutrino
  masses, the Higgs boson can decay mainly to the invisible Goldstone
  boson associated to the spontaneous breaking of lepton number. This
  requires a gauge singlet superfield coupling to the electroweak
  doublet Higgses, as in the Next to Minimal Supersymmetric Standard Model 
  (NMSSM) scenario for solving the
  $\mu$-problem.  The search for invisibly decaying Higgs bosons
  should be taken into account in the planning of future accelerators,
  such as the Large Hadron Collider and the Next Linear Collider.

\end{abstract}  


\maketitle

\section{Introduction
        \label{Introduction}}
      
      Understanding the origin of mass is the main open puzzle in
      particle physics today.  In the Standard Model all masses arise
      as a result of the spontaneous breaking of the \21 gauge
      symmetry.  This implies the existence of an elementary Higgs
      boson, not yet found. Stabilizing the mass of the Higgs most
      likely requires new physics and supersymmetry has so far been
      the leading contender. Another aspect of this problem is the
      smallness of neutrino masses.
      Despite the tremendous effort that has led to the discovery of
      neutrino mass~\cite{fukuda:1998mi,ahmad:2002jz,eguchi:2002dm}
      the mechanism of neutrino mass generation will remain open for
      years to come (a detailed analysis of the three--neutrino
      oscillation parameters can be found in ~\cite{Maltoni:2004ei}).
      The most popular mechanism to generate neutrino masses is the
      seesaw
      mechanism~\cite{gell-mann:1980vs,yanagida:1979,mohapatra:1981yp,chikashige:1981ui,schechter:1980gr}.
      Although the seesaw fits naturally in SO(10) unification models,
      we currently have no clear hints that uniquely point towards any
      unification scheme.
      Therefore it may well be that neutrino masses arise from
      garden--variety physics having nothing to do with unification,
      such as certain seesaw variants~\cite{mohapatra:1986bd}, and models 
      with radiative generation~\cite{zee:1980ai,babu:1988ki}.
      In such models the physics of neutrino mass would then be
      characterized by much lower scales~\cite{valle:1991pk},
      potentially affecting the decay properties of the Higgs boson.
      This is especially so if neutrino masses arise due to the
      spontaneous violation of ungauged lepton number. In this broad
      class of models the Higgs boson will have an important decay
      channel into the singlet Goldstone boson (called majoron)
      associated to lepton number violation~\cite{Joshipura:1993hp},
      \begin{equation}
        \label{eq:HJJ}
        h \to J J\,.
      \end{equation}
         
      Here we focus on the specific case of low-energy supersymmetry
      with spontaneous violation of R--parity, as the origin of
      neutrino mass. R--parity is defined as $R_p
      = (-1)^{3B+L+2S}$ with $S$, $B$, $L$ denoting spin, baryon and
      lepton numbers, respectively~\cite{aulakh:1982yn}.
      In this model R--parity violation takes place ``a la Higgs'', i.e.,
      spontaneously, due to non-zero sneutrino vacuum expectation values
      (vevs)~\cite{masiero:1990uj,romao:1992vu,shiraishi:1993di}.
      In this case one of the neutral CP-odd scalars is
      identified with the majoron. In
      contrast with the seesaw majoron, ours is characterized by a
      small scale (TeV-like) and carries only one unit of lepton
      number.
      This scheme leads to the bilinear R--parity violation
      model, the simplest
      effective description of R-parity violation~\cite{diaz:1998xc} (for 
      calculations including also trilinear terms see, 
      for example \cite{chun:1999bq,abada:2001zh}). The model not only
      accounts for the observed pattern of neutrino masses and
      mixing~\cite{Diaz:2003as,hirsch:2000ef,romao:1999up,Hirsch:2004he}, 
      but also
      makes predictions for the decay branching ratios of the lightest
      supersymmetric
      particle~\cite{hirsch:2002ys,porod:2000hv,restrepo:2001me,Hirsch:2003fe}
      from the current measurements of neutrino mixing
      angles~\cite{Maltoni:2004ei}.
      
      In previous studies~\cite{deCampos:1995av} it was noted that the
      spontaneously broken R--parity (SBRP) model leads to the
      possibility of invisibly decaying Higgs bosons, provided there
      is an \21 singlet superfield $\Phi$ coupling to the electroweak
      doublet Higgses, the same that appears in the NMSSM.
      
      In this paper we reanalyse this issue taking into account the
      small masses indicated by current neutrino oscillation
      data~\cite{Maltoni:2004ei}~\footnote{Ref.~\cite{deCampos:1995av}
        assumed MeV-scale for the heaviest neutrino mass, inconsistent
        with the atmospheric data which points towards $m_\nu \sim
        0.05$~eV.}.  We focus on the lowest-lying neutral CP-even
      scalar boson of the model.  We show explicitly that the presence
      of the \21 singlet superfield $\Phi$ plays a triple role: (i) it
      gives a model where neutrino masses are obtained from first
      principles without any type of fine-tuning, even when radiative
      corrections are negligible, (ii) it solves the $\mu$-problem ``a
      la NMSSM''~\footnote{Provided domain walls are either eliminated
        by imposing a ${\cal {Z}}_2$ $R$-symmetry on the
        non-renormalizable operators~\cite{Panagiotakopoulos:1998yw},
        or that they are simply inflated away.}, and (iii) it makes
      the invisible Higgs boson decay in Eq.~(\ref{eq:HJJ})
      potentially the most important mode of Higgs boson decay. The
      latter is remarkable, given the smallness of neutrino masses
      required to fit current neutrino oscillation data.  We also
      verify that the production of such Higgs boson in $e^+e^-$
      annihilation can be as large as that characterizing the standard
      case, and that therefore this situation should be taken as part
      of the agenda of future accelerators probing the mechanism of
      mass generation.

\section{Model with spontaneously broken R parity
  \label{Model with spontaneous broken R parity}}

The most general superpotential terms involving the Minimal 
Supersymmetric Standard Model (MSSM) superfields
in the presence of the \21 singlet superfields $({\widehat \nu^c}_i,
\widehat{S}_i,{\widehat \Phi})$ 
carrying a conserved lepton number assigned as $(-1, 1,0)$, respectively 
is given as~\cite{romao:1992zx} 
\begin{eqnarray} \nonumber
{\cal W} &\hskip-4mm=\hskip-4mm& \varepsilon_{ab}\Big(
h_U^{ij}\widehat Q_i^a\widehat U_j\widehat H_u^b 
+h_D^{ij}\widehat Q_i^b\widehat D_j\widehat H_d^a 
+h_E^{ij}\widehat L_i^b\widehat E_j\widehat H_d^a 
+h_{\nu}^{ij}\widehat L_i^a\widehat \nu^c_j\widehat H_u^b 
\!- {\hat \mu}\widehat H_d^a\widehat H_u^b 
\!- (h_0 \widehat H_d^a\widehat H_u^b +\delta^2)\widehat\Phi \Big)  \\
& &+\hskip 5mm   h^{ij} \widehat\Phi \widehat\nu^c_i\widehat S_j +
M_{R}^{ij}\widehat \nu^c_i\widehat S_j 
+ \frac{1}{2}M_{\Phi} \widehat\Phi^2 +\frac{\lambda}{3!} \widehat\Phi^3
\label{eq:Wsuppot} 
\end{eqnarray}
The first three terms together with the $ {\hat \mu}$ term define the
R-parity conserving MSSM, the terms in the last row only involve the
\21 singlet superfields 
$({\widehat \nu^c}_i,\widehat{S}_i,{\widehat \Phi})$~\footnote{The term
  linear in $\Phi$ has been included in the first row as it is
  relevant in electroweak breaking.}, while the remaining terms couple
the singlets to the MSSM fields. We stress the importance of the
Dirac-Yukawa term which connects the right-handed neutrino superfields
to the lepton doublet superfields, thus fixing lepton number.

\subsection{Spontaneous Symmetry Breaking}
\label{sec:spontaneous-symmetry-breaking}

The presence of singlets in the model is essential in order to drive
the spontaneous violation of R parity and electroweak symmetries in a
phenomenologically consistent way.  Like all other Yukawa couplings
$h_U, h_D, h_E$ we assume that $h_{\nu}$ is an arbitrary non-symmetric
complex matrix in generation space. For technical simplicity we take
the simplest case with just one pair of lepton--number--carrying \21
singlet superfields, $\widehat\nu^c$
and $\widehat S$, in order to avoid
inessential complication. This in turn implies, $ h_{ij} 
\to h$ and $ h_{\nu}^{ij} \to h_{\nu}^{i}$. 

The full scalar potential along neutral directions is given by
\begin{eqnarray}
V_{total}  &=&
|h \Phi \tilde{S} + h_{\nu}^{i} \tilde{\nu}_i H_u  + M_R \tilde{S}|^2 +
|h_0 \Phi H_u + \hat{\mu} H_u|^2 + |h \Phi \tilde{\nu^c}+ M_R
\tilde{\nu^c}|^2 
\label{scalarpot}
\\\nonumber&&
+|- h_0 \Phi H_d  - \hat{\mu} H_d +
h_{\nu}^{i} \tilde{\nu}_i \tilde{\nu^c} |^2+
|- h_0 H_u H_d + h \tilde{\nu^c} \tilde{S} - \delta^2 +M_{\Phi} \Phi 
+\frac{\lambda}{2} \Phi^2|^2 \\\nonumber&&
+\sum_{i=1}^3|h_{\nu}^{i} \tilde{\nu^c} H_u|^2
+ \Big[  A_h h \Phi \tilde{\nu^c} \tilde{S}
- A_{h_0} h_0 \Phi H_u H_d 
+ A_{h_{\nu }}  h_{\nu}^{i}\tilde{\nu}_i H_u \tilde{\nu^c} 
- B \hat{\mu} H_u H_d  \\\nonumber&&
- C_{\delta} \delta^2 \Phi + B_{M_R} M_R \tilde{\nu^c} \tilde{S} 
+ \frac{1}{2} B_{M_{\Phi}} M_{\Phi} \Phi^2 
+ \frac{1}{3!} A_{\lambda} \lambda \Phi^3 
 + h.c. \Big]\\\nonumber&&
+ \sum_{\alpha} \tilde{m}_{\alpha}^2 |z_{\alpha}|^2
+ \frac{1}{8} (g^2 + {g'}^2) 
\Big( |H_u|^2 - |H_d|^2 - \sum_{i=1}^3 |\tilde{\nu}_i|^2\Big)^2,
\end{eqnarray}
where $z_{\alpha}$ denotes any neutral scalar field in the theory.

The pattern of spontaneous symmetry breaking of both electroweak and R
parity symmetries works in a very simple way. The spontaneous breaking
of R parity is driven by nonzero vevs for the scalar neutrinos. The
scale characterizing R parity breaking is set by the isosinglet vevs
\begin{equation}
\vev{\tilde{\nu^c}} = \frac{v_R}{\sqrt{2}},\quad
\langle\tilde{S}\rangle=\frac{v_S}{\sqrt{2}},
\label{eq:rpv}
\end{equation}
and 
\begin{equation}
\vev{\Phi} = \frac{v_{\Phi}}{\sqrt{2}}.
\label{eq:rpphi}
\end{equation}
We also have  very small left-handed sneutrino vacuum expectation
values
\begin{equation}
\vev{\tilde{\nu}_{Li}} = \frac{v_{Li}}{\sqrt{2}}\,.
\label{vl}
\end{equation}
The spontaneous breaking of R--parity also entails the spontaneous
violation of total lepton number. This implies that one of the neutral 
CP--odd scalars, which we call majoron, and which is given by 
the imaginary part of
\begin{equation}
\frac{\sum_i v_{Li}^2}{Vv^2} (v_u H_u - v_d H_d) +
\sum_i \frac{v_{Li}}{V} \tilde{\nu_{i}} 
+\frac{v_S}{V} S
-\frac{v_R}{V} \tilde{\nu^c}
\label{maj}
\end{equation}
remains massless, as it is the Nambu-Goldstone boson associated to the
breaking of lepton number. Note that this majoron is quite different
from the one that emerges in the seesaw majoron model, as it is
characterized by a different lepton number (one unit instead of two)
and by a different scale, determined by the combination $V =
\sqrt{v_R^2 + v_S^2} \sim$~TeV .  Note that Eq.~(\ref{eq:rpv}) is the
origin of lepton number violation in this model and plays a crucial
role in determining the neutrino masses. 

On the other hand, electroweak breaking is driven by the isodoublet vevs 
$\vev{{H_u}} = \frac{v_{u}}{\sqrt{2}}$ and 
$\vev{{H_d}}= \frac{v_{d}}{\sqrt{2}}$, with the combination 
$v^2 = v_u^2 + v_d^2 + \sum_i v_{L i}^2$ fixed by the W mass
\begin{equation}
m_W^2 = \frac{g^2 v^2}{4},
\label{mw}
\end{equation}
while the ratio of isodoublet vevs yields
\begin{equation}
\tan \beta = \frac{v_u}{v_d}.
\label{beta}
\end{equation}
This basically recovers the standard tree level spontaneous breaking
of the electroweak symmetry in the
MSSM~\cite{Barbieri:1982eh}~\footnote{ We have verified explicitly,
  however, that radiative electroweak breaking may also occur.  }.

\subsection{Neutrino masses}
\label{sec:neutrino-masses}

Since neutrino masses are so much smaller than all other fermion mass
terms in the model, once can find the effective neutrino mass matrix
in a seesaw--type approximation. From the full neutral fermion mass
matrix, see Eq.~(\ref{eq:mass}), one calculates the effective $3\times
3$ neutrino mass matrix $(\mathbf{m_{\nu\nu}^{\rm eff}})$ as
\begin{equation}
\mathbf{m_{\nu\nu}^{\rm eff}} = -\mathbf{M_D^T}\mathbf{M_H^{-1}}\mathbf{M_D},
\label{eq:Seesaw}
\end{equation}
where $\mathbf{M_H}$ is the $7\times 7$ matrix of all other neutral
fermion states, see Eq.~(\ref{eq:mass}), and the $3\times 7$ matrix
$\mathbf{m^T_{\chi^0\nu}}$ is given as
\begin{equation}
\mathbf{M_D^T}=
\left(\begin{array}{llll}
\mathbf{m^T_{\chi^0\nu}} & \mathbf{m_{D}} & 
\mathbf{0} & \mathbf{0} 
\end{array} \right),
\label{eq:massDSim}
\end{equation}
where the matrices $\mathbf{m^T_{\chi^0\nu}}$ and $\mathbf{m_{D}}$ are
given in Eqs.~(\ref{eq:mrpv}) and (\ref{eq:mD}). The inverse of
$\mathbf{M_H}$ is too long to be given explicitly here.

After some algebraic manipulation, the effective neutrino mass 
matrix can be cast into a very simple form 
\begin{equation}
(\mathbf{m_{\nu\nu}^{\rm eff}})_{ij} = a \Lambda_i \Lambda_j + 
     b (\epsilon_i \Lambda_j + \epsilon_j \Lambda_i) +
     c \epsilon_i \epsilon_j.
\label{eq:eff}
\end{equation}
where one can define the effective bilinear R--parity violating
parameters $\epsilon_{i}$ and $\Lambda_i$ as
\begin{equation}
  \label{eq:eps}
\epsilon_{i} = h_{\nu}^{i}\, \frac{v_R}{\sqrt{2}}  
\end{equation}
and
\begin{equation}
\Lambda_i = \epsilon_i v_d + \mu v_{L_i}
\label{eq:deflam0}
\end{equation}
Here the parameter $\mu$ is
\begin{equation}
\mu = \hat{\mu} + h_0 \frac{v_{\Phi}}{\sqrt{2}}
\label{eq:defmu},
\end{equation}
while the coefficients appearing in Eq.~(\ref{eq:eff}) are given by  
\begin{equation}
a= \frac{1}{4\mu \rm{Det}(\mathbf{M_H})}\Big(m_{\gamma}\widehat{M}_R
(-h^2 v_{R} v_{S}\mu + \widehat{M}_{\Phi}\widehat{M}_R\mu 
+h_0^2\widehat{M}_Rv_dv_u) \Big)
\label{eq:co_a}
\end{equation}
\begin{equation}
b= \frac{1}{8\mu \rm{Det}(\mathbf{M_H})}\Big(h_0 m_{\gamma}\widehat{M}_R
(h_0 \widehat{M}_R + h\mu)v_u(v_u^2-v_d^2)\Big)
\label{eq:co_b}
\end{equation}
\begin{equation}
c= \frac{1}{4\mu \rm{Det}(\mathbf{M_H})}\Big(
(h_0 \widehat{M}_R + h\mu)^2v_u^2 (2 M_1M_2\mu -m_{\gamma}v_dv_u)\Big)
\label{eq:co_c}
\end{equation}
and $\rm{Det}(\mathbf{M_H})$ is given as

\begin{eqnarray}\nonumber
\rm{Det}(\mathbf{M_H})&= & \frac{1}{8}\widehat{M}_R
\Big\{8M_1M_2\mu(\widehat{M}_{\Phi}\widehat{M}_R\mu-
h^2\mu v_Rv_S+h_0^2\widehat{M}_R v_dv_u) \\ 
&-&m_{\gamma}\Big(4\mu v_d(\widehat{M}_{\Phi}\widehat{M}_R-h^2v_Rv_S)
v_u +h_0^2\widehat{M}_R(v_d^2+v_u^2)^2\Big)\Big\}
\label{eq:det}
\end{eqnarray}
Note that $\widehat{M}_R$ and $\widehat{M}_{\Phi}$ above are 
defined as 
\begin{eqnarray}
\widehat{M}_R=M_R + h \frac{v_{\Phi}}{\sqrt{2}},\quad
\widehat{M}_{\Phi}=M_{\Phi} + \lambda \frac{v_{\Phi}}{\sqrt{2}}.
\label{eq:effmr}
\end{eqnarray}
The ``photino'' mass parameter is defined as $m_{\gamma} = g^2M_1
+g'^2 M_2$.

Eq.~(\ref{eq:eff}) resembles very closely the corresponding expression 
for the explicit bilinear R-parity breaking model~\cite{diaz:1998xc,chun:1999bq,abada:2001zh,Diaz:2003as,hirsch:2000ef,romao:1999up},  
once the dominant 1-loop corrections are taken into account. 
Note that the tree-level result of the explicit bilinear model 
can be recovered in the limit 
$\widehat{M}_R,\widehat{M}_{\Phi} \to \infty$. 
In this limit the coefficients $b$ and $c$ go to zero, while 

\begin{equation}
a = \frac{m_{\gamma}}{4{\rm Det}(\mathbf{M_{\chi^0}})}
\label{TreeLevelBil}
\end{equation}
In this limit only one non-zero neutrino mass remains. Whether the
1-loop corrections or the contribution from the singlet fields are
more important in determining the neutrino masses depends essentially
on the relative size of the coefficient $c$ in Eq.~(\ref{eq:eff})
compared to the corresponding 1-loop coefficient. Both extremes can be
realized in our model. We note, however, that as discussed below large
branching ratios of the Higgs into invisible final states require
sizeable values of $h$ and $h_0$ (as well as singlets not being too
heavy). For such choices of parameters we have found that the
``singlino'' contribution to Eq.~(\ref{eq:eff}) is usually much more
important than the 1-loop corrections to the neutrino masses.

Note also that the model does not predict whether the atmospheric 
(solar) mass scale is mainly due to the first (third) term in 
Eq.~(\ref{eq:eff}) or vice versa. We have checked numerically that 
both possibilities can be realized and ``good'' points (in the sense 
of being appropriate for neutrino physics) can be found easily 
in either case.

\subsection{Scalar Mass Matrices}
\label{sec:scalar-mass-matrices}

With the above choices and definitions we can obtain the neutral scalar 
boson mass matrices as in Ref.~\cite{romao:1992vu} by evaluating the second
derivatives of the scalar potential in Eq.~(\ref{scalarpot}) at the
minimum. This results in $8 \times 8$ mass matrices for the real
and imaginary parts of the neutral scalars~\footnote{As already
  mentioned we assume, for technical simplicity that we have just one
  pair of lepton--number--carrying \21 singlet superfields.}. 
We have checked, in particular, that in the CP-odd sector
we find both the Goldstone ``eaten'' by the $Z^0$ as well as the
Goldstone boson corresponding to the spontaneous breaking of R-parity,
namely the majoron, Eq.~(\ref{maj}).
In the basis
$A'_0=(H_d^{0 I},H_u^{0 I},\tilde{\nu}^{1 I},\tilde{\nu}^{2 I},
\tilde{\nu}^{3 I},\Phi^I,\tilde{S}^I,\tilde{\nu}^{c I})$ 
these fields are given as,
\begin{eqnarray}
  \label{eq:3a}
  G_0&=&(N_0\, v_d,-N_0\, v_u,N_0\,v_{L1},N_0\, v_{L2},N_0\, v_{L3},0,0,0)
  \nonumber \\[+2mm]
  J&=&N_4 (-N_1 v_d,N_1 v_u, N_2 v_{L1}, N_2 v_{L2}, N_2 v_{L3},0,
  N_3 v_S,-N_3 v_R)
\end{eqnarray}
where the normalization constants $N_i$ are given as
\begin{eqnarray}
  \label{eq:4a}
  N_0&=&\frac{1}{\sqrt{v_d^2+v_u^2+v_{L1}^2 + v_{L2}^2 + v_{L3}^2}}
\nonumber \\[+2mm]
  N_1&=&v_{L1}^2 + v_{L2}^2 + v_{L3}^2\nonumber \\[+2mm]
  N_2&=&v_d^2 + v_u^2\nonumber \\[+2mm]
  N_3&=&N_1 + N_2 \nonumber \\[+2mm]
  N_4&=&\frac{1}{\sqrt{N_1^2 N_2 + N_2^2 N_1 +N_3^2(v_R^2+v_S^2)}}
\end{eqnarray}
and can easily be checked to be orthogonal, i.~e. they satisfy $G_0
\cdot J=0$. 


In order to study the phenomenology of the scalar sector we need some
information about the parameters of the SBRP model. Broadly speaking
there are four types of parameters that are to a large extent
undetermined. First there are Yukawa couplings, like $h$, $h_0$ and 
$\lambda$. In contrast to $h_U$, $h_D$ and $h_E$ these are not fixed 
by fermion masses.
Then there are MSSM parameters such as $\tan\beta$, the effective
Higgsino mixing parameter $\mu$, the supersymmetry breaking scalar
mass parameters $m_0$ and $A_0$.  These are partially restricted by
negative collider searches for supersymmetric
particles~\cite{abreu:2000mm}. Then there are singlet sector mass
parameters, such as $M_R$, $M_\Phi$ and $\delta^2$.  Finally there is
the important Yukawa coupling $h_\nu$, which determines the strength
of effective R--parity breaking parameters, through
Eq.~(\ref{eq:eps}). This is constrained by neutrino oscillation data.
In Section~\ref{sec:numerical-results} we will discuss our strategy to
choose the parameters in such a way that the results can be easily
interpreted. We will also show there that a fully cubic
superpotential, without any mass scale parameter such as the $
\hat{\mu} H_u H_d$ term, also leads to a realistic
model~\cite{Romao:1997xf} consistent with neutrino oscillation data.
Before that, however, we consider the corresponding Higgs boson
phenomenology, focusing on Higgs boson production and decays, and
stressing the potentially large invisible decay branching ratio.

\section{Higgs boson production and decays}
\label{sec:higgs-prod-decays}

Supersymmetric Higgs bosons can be produced at the $e^+ e^-$ collider
through their couplings to $Z$, via the so--called Bjorken process.
In our SBRP model there are 8 neutral CP--even states $H_i$ and 6
neutral CP--odd Higgs bosons $A_i$, in addition to the majoron $J$.
One must diagonalize the scalar boson mass matrix in order to find the
coupling of the massive scalars to the $Z$.  The Lagrangean
is
\begin{equation}
\label{HZZ1}
{\cal L}_{HZZ}
=\sum_{i=1}^8 (\sqrt 2 G_F)^{1/2} M_Z^2 Z_{\mu}Z^{\mu} \eta_{i} H_i
\end{equation}
with each $\eta_i$ given as a weighted combination of the five \21
doublet scalars,
\begin{equation}
  \label{eq:1}
  \eta_i= \frac{v_d}{v} R^S_{i 1} + \frac{v_u}{v} R^S_{i 2} + 
\sum_{j=1}^3 \frac{v_{Lj}}{v} R^S_{i j+2} 
\end{equation}
where $R^S_{ij}$ is the $8 \times 8$ rotation matrix for the CP-even
scalars. Note that we leave the discussion of the CP--odd scalars for
elsewhere.
Moreover, here we focus mainly on the production of the lightest
CP--even supersymmetric Higgs boson $h \equiv H_1$. The main
difference between the production of this state and the lightest
CP--even Higgs boson of the MSSM is the fact that ours contains an
admixture of the \21 singlet scalar fields ${\tilde \nu}^c$ and
$\tilde{S}$, and its coupling to the $Z$ is correspondingly reduced by
a factor
\begin{equation}
  \label{eq:eta}
   \eta \equiv \eta_1 \leq 1
\end{equation}
in comparison with the Standard Model case~\footnote{For the MSSM we
  have a reduction given by $\eta=\frac{v_d}{v} R^S_{1 1} +
  \frac{v_u}{v} R^S_{1 2}= \sin(\beta-\alpha)$ in the usual
  notation.}.  When the lightest CP--even Higgs boson is mainly
singlet its production cross section in $e^+ e^-$ annihilation will be
suppressed.

We now turn to the lightest Higgs boson decays. Given that other MSSM
decay modes are less important, we are particularly interested here in
the ratio
\begin{equation}
  \label{eq:ratio}
  R_{Jb}=\frac{\Gamma(h \to JJ)}{\Gamma(h \to b \bar{b})}
\end{equation}
of the invisible decay to the Standard Model decay into b-jets. For
this we have to look separately at the decay widths,
\begin{equation}
  \label{eq:JJ}
  \Gamma(h\to  JJ)=\frac{g_{hJJ}^2}{32\pi m_h}
\end{equation}
and
\begin{equation}
  \label{eq:bb}
  \Gamma(h\to  b \bar{b})=\frac{3 G_F \sqrt{2}}{8\pi\cos^2\beta}\,
  \left(R^S_{11}\right)^2\, m_h\,  m_b^2 \left[ 1-4
  \left(\frac{m_b}{m_h}\right)^2 \right]^{3/2}
\end{equation}
From these expressions we see that $\Gamma(h \to b \bar{b})$ will be
small if the component of the lightest Higgs boson along $H_d^0$ is
small. On the other hand the magnitude of $\Gamma(h \to JJ)$ will
depend on the $g_{hJJ}$ coupling. This is in general given by a
complicated expression, but for the situation that we are considering
here with
\begin{equation}
  \label{eq:vi}
  v_{Li} \ll v_d,v_u \ll v_R,v_S
\end{equation}
we have to a very good approximation
\begin{equation}
  \label{eq:14a}
  J\simeq (0,0,0,0,0,0,\frac{v_S}{V},-\frac{v_R}{V})
\end{equation}
where $  V^2=v_S^2+v_R^2$.
Under this approximation we can write the coupling $g'_{i}$ for the
vertex $h'_i JJ$ of the Majoron with the \textit{unrotated} Higgs
boson $h'_i$, in the following form
\begin{eqnarray}
  \label{eq:16}
  g'_1&=&h h_0 v_u \frac{v_S v_R}{V^2}\nonumber \\[+1mm]
  g'_2&=&h h_0 v_d \frac{v_S v_R}{V^2} -\frac{2 v_u}{V^2}\,
  \sum_{j=1}^3 \epsilon_j^2
\nonumber \\[+1mm]
  g'_i&=&-\frac{2 \epsilon_{i-2}}{V^2}\sum_{j=1}^3 \epsilon_j v_{Lj}
  \quad (i=3,4,5)
\nonumber \\[+1mm]
  g'_6&=& -\sqrt{2}\, h \left(A_h +   \widehat{M}_{\Phi} \right)\,
\frac{v_S v_R}{V^2} -\sqrt{2}\, h\, \widehat{M}_R \nonumber \\[+1mm]
  g'_7&=&-h^2 \frac{v_S v_R^2}{V^2}\nonumber \\[+1mm]
  g'_8&=&-h^2 \frac{v_S^2 v_R}{V^2}
\end{eqnarray}
where $\widehat{M}_R$ and $\widehat{M}_{\Phi}$ have been defined in 
Eq.~(\ref{eq:effmr}).

From these expressions we conclude that $g_{hJJ}$ can be large in two
situations. The first is, of course, if the lightest Higgs boson is
mainly a combination of the $\tilde{\nu^c}$ and $\tilde{S}$ fields.
In this case not only $g_{hJJ}$ will be large, but also
$\Gamma(h\to b{\bar b})$ will be small suppressing $h \to b
\bar{b}$. Unfortunately the production would be suppressed, as
singlets do not couple to the $Z$.
The phenomenologically novel and interesting situation is when $h$ and
$h_0$ are large. In this case the Higgs boson behaves as the lightest
MSSM Higgs boson (with moderately reduced production cross section)
but with a large branching to the invisible channel $h \to JJ$.

The sensitivities of LEP experiments to the invisible channel $h \to
JJ$ have been discussed since long
ago~\cite{lopez-fernandez:1993tk,deCampos:1997bg} and the current
status has been presented in Ref.~\cite{Abdallah:2003ry}.  In order to
evaluate the experimental sensitivities to the parameters of the model
we must take into account both the production as well as Higgs decays.

\section{Numerical results}
\label{sec:numerical-results}

In this section we discuss the numerical results on the invisible
decay of the Higgs boson in our model. We start with a brief
discussion of the SBRP parameters.

Unknown parameters of the spontaneous R-parity breaking model fall
into three different groups. First, there are the MSSM parameters,
mainly the unknown soft SUSY breaking terms. The second group of
parameters are the $\epsilon_i$ and left-handed sneutrino vevs
$v_{L_i}$.  We trade the latter for the parameters $\Lambda_i$ using
Eq.~(\ref{eq:deflam0}).  These six parameters occur also in the
explicit bilinear model.  And, finally, there are the parameters of
the singlet sector, namely singlet vevs $v_R$, $v_S$ and $v_{\Phi}$,
Yukawa couplings $h$, $h_0$ and $\lambda$ and the singlet mass terms
$M_R$, $M_{\Phi}$, $\delta^2$, as well as the corresponding soft
terms.

We have checked by a rather generous scan that the results presented 
below qualitatively do not depend on the choice of MSSM parameters, 
as expected. Thus, for definiteness we will fix the MSSM parameters 
in the following to the SPS1a benchmark point \cite{Allanach:2002nj}, 
defined by 

\begin{equation}
m_0 = 100 \rm{GeV} \hskip5mm m_{1/2} = 250 \rm{GeV} \hskip5mm \tan\beta=10 
\hskip5mm A_0 = -100 \rm{GeV}  \hskip5mm \mu <0
\label{sps1a}
\end{equation}
We have run down this set of parameters to the electro-weak scale using 
the program package SPheno \cite{SPheno}. We stress again that different 
choices of MSSM parameters will not lead to qualitatively different 
results.

\subsection{General case}
\label{sec:general-case}

We first consider the general model defined by the superpotential in
Eq.~(\ref{eq:Wsuppot}) reduced to one generation of $\nu^c$ and $S$
fields.  For the singlet parameters we choose as a starting point
$v_R= v_S = v_{\Phi} = -150$ GeV and $M_R = - M_{\Phi} = \delta =
10^3$ GeV, as well as $h=0.8$, $h_0 = -0.15$ and $\lambda=0.1$. We
have tried other values of parameters and obtained qualitatively
similar results to the ones discussed below.

The explicit bilinear parameters are then fixed approximately such
that neutrino masses and mixing angles~\cite{Maltoni:2004ei} are in
agreement with experimental
data~\cite{fukuda:1998mi,eguchi:2002dm,ahmad:2002jz}. Slightly
different values of parameters are found, depending on whether the
first or the third term in Eq.~ (\ref{eq:eff}) is responsible for the
atmospheric neutrino mass scale. Both possibilities lead to very
similar results for the invisible decay of the Higgs.  This can be
understood quite easily.  The ratio of the atmospheric and solar
neutrino mass scale is only of the order of ($4-7$) \footnote{In a
  hierarchical model, such as the one discussed here, the square roots
  of the $\Delta m_{ij}^2$ are approximately equal to the larger
  mass.} and the changes in parameters $\vec \Lambda$ and $\vec
\epsilon$ are only of the order of the square root of this number.
Such a small change can always be compensated by a slight adjustment
of other parameters, leading to the same (or very similar) final
result.

\begin{figure}[htbp]
\begin{center}
\vspace{5mm}
\includegraphics[width=75mm,height=5cm]{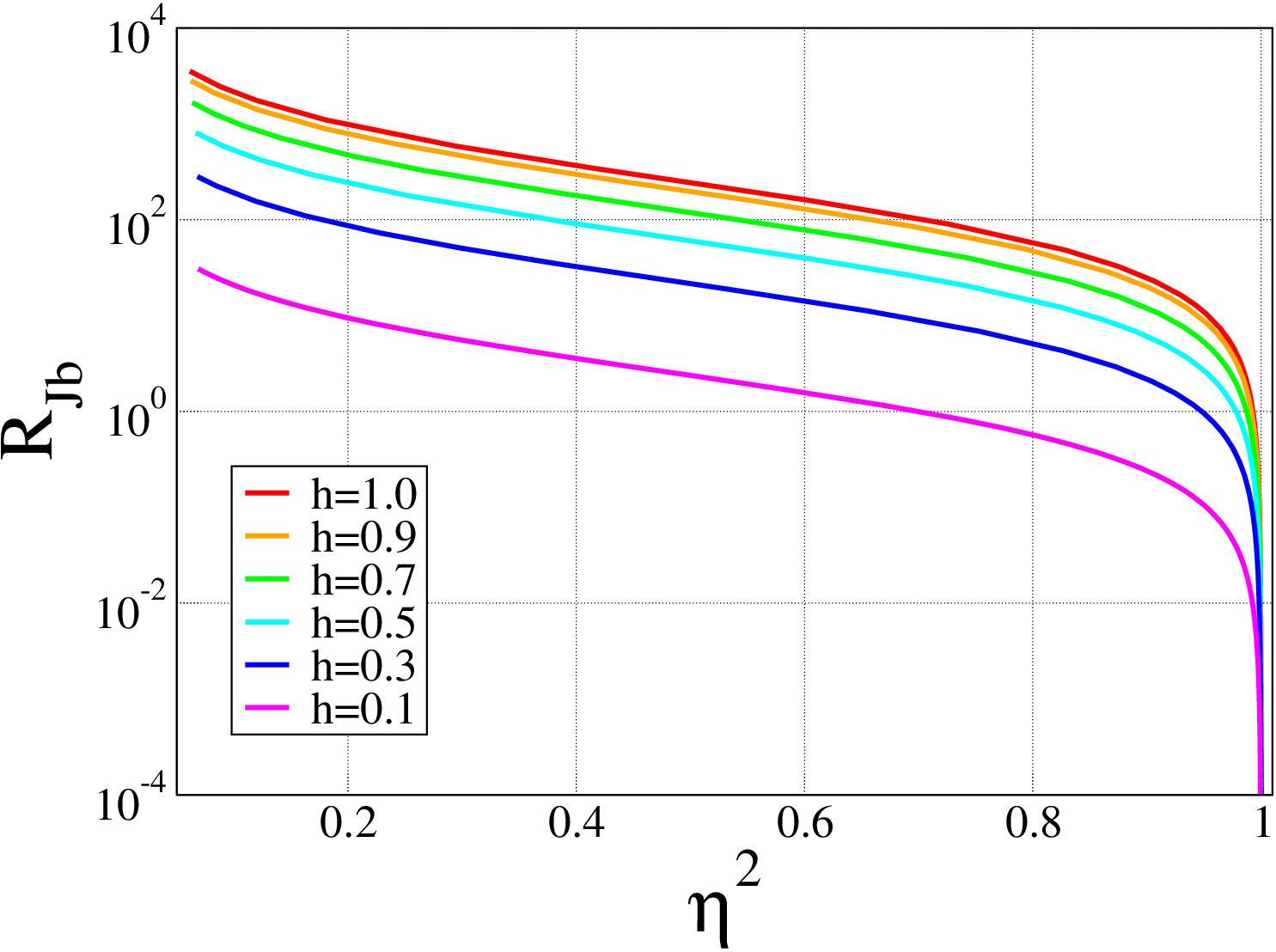}
\includegraphics[width=75mm,height=5cm]{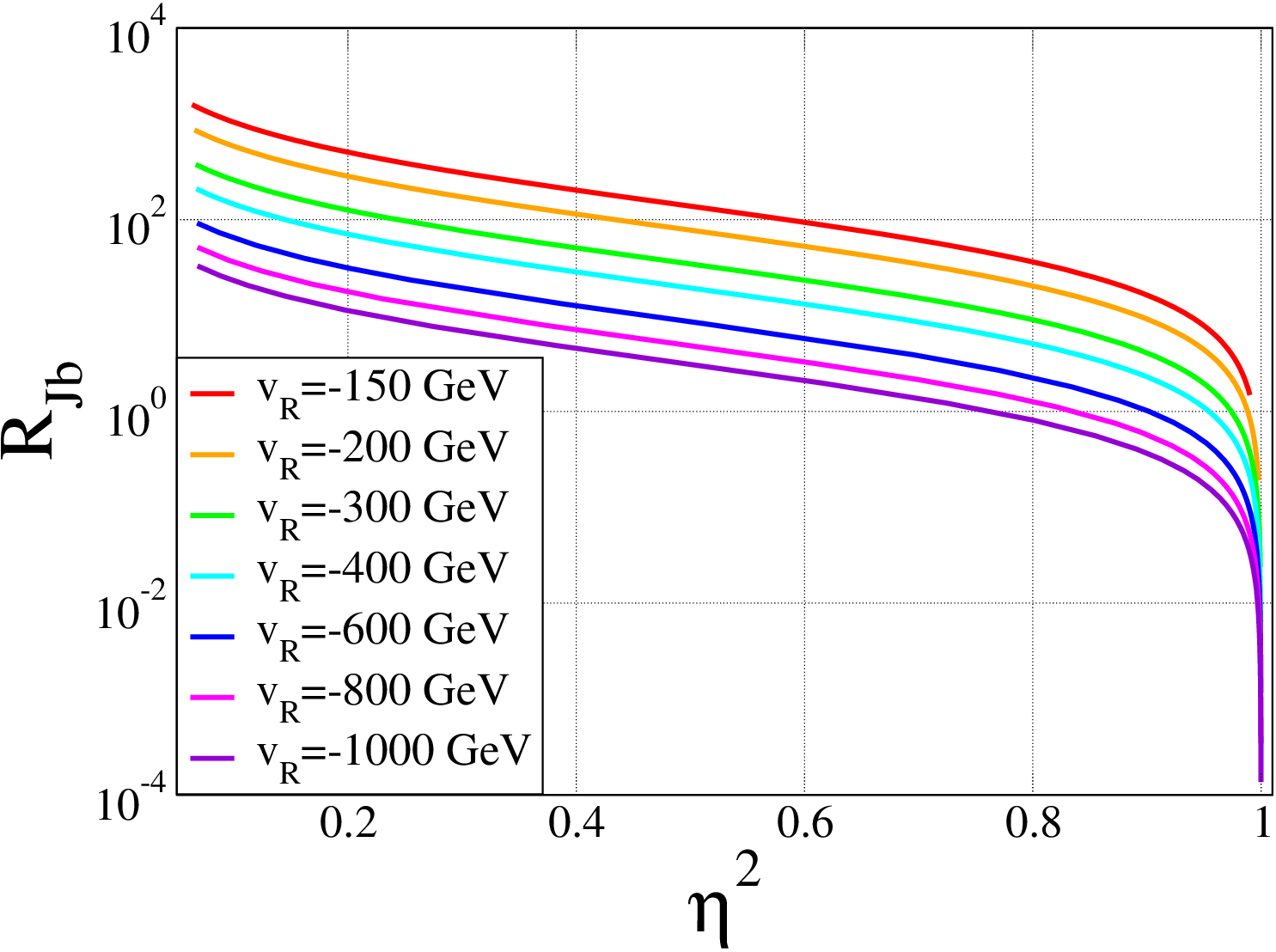}
\end{center}
\vspace{-5mm}
\caption{Ratio $R_{Jb}$, defined in Eq.~(\ref{eq:ratio}), as function 
  of $\eta^2$.  a) to the left, for different values of the
  parameter $h$, from top to bottom: $h=1,\, 0.9,\, 0.7,\, 0.5,\, 0.3,\, 0.1$. 
  b) to the right, for different values of the
  parameter $v_R=v_S$: $-v_R= 150,\, 200,\, 300,\, 400,\, 600,\, 800,\, 1000$ GeV. 
  The plots show explicitly that $R_{Jb}>1$ is possible even for 
  $\eta \simeq 1$. This is the main result of the current paper.}
\label{fig:plot54}
\end{figure}

After having defined our ``preferred'' choice of parameters in the
following we will vary one unknown parameter at a time. We now turn to
a discussion of the results. In Fig. (\ref{fig:plot54}) we show the
ratio $R_{Jb}$ as a function of $\eta^2$ for different choices of $h$ (left) 
and for different choices of $v_R$ (right) and all other parameters fixed. 
Larger values of $R_{Jb}$ are found for smaller values of $\eta$, as 
expected. However, one sees explicitly that even for values of 
$\eta \simeq 1$, $R_{Jb}$ can be larger than $1$. This means that the 
lightest Higgs can decay mainly invisibly, even when the cross section 
for its production is essentially equal to the usual (MSSM) doublet Higgs 
boson cross section. This is the main result of this work.

\begin{figure}[htbp]
\vspace{5mm}
\begin{center}
\includegraphics[width=75mm,height=5cm]{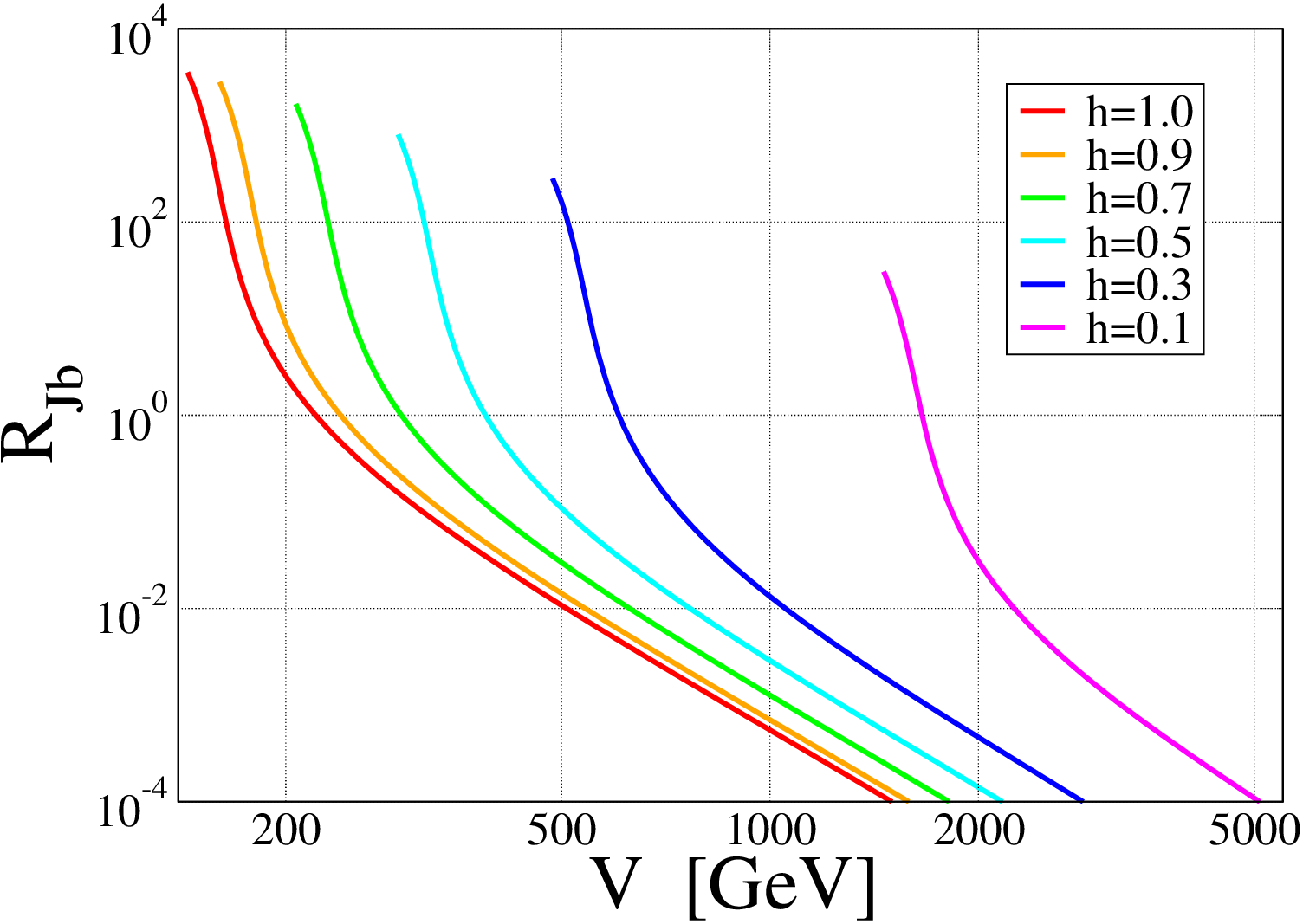}
\includegraphics[width=75mm,height=5cm]{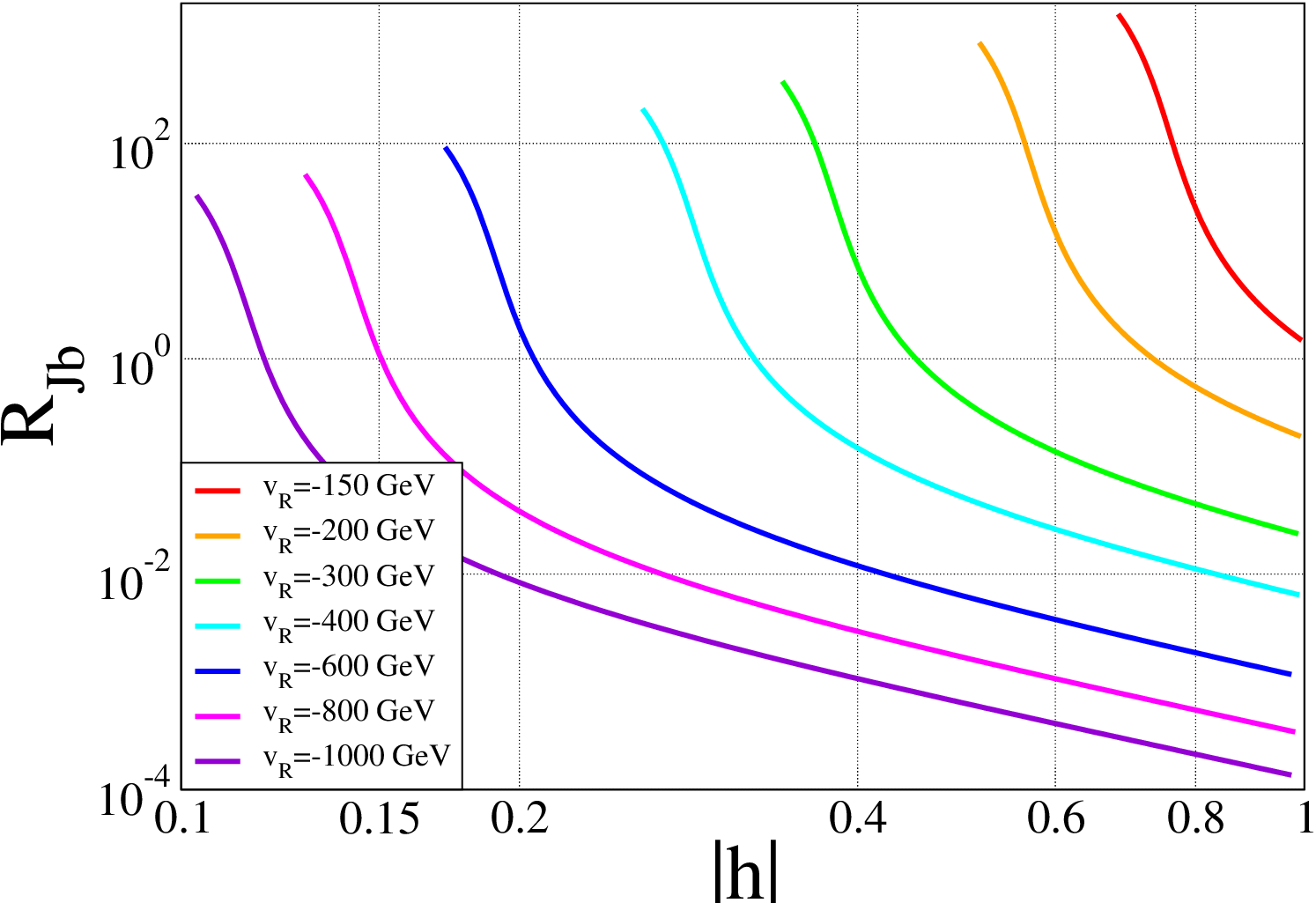}
\end{center}
\vspace{-5mm}
\caption{Ratio $R_{Jb}$, defined in Eq.~(\ref{eq:ratio}), as function 
  of $V$ (left) and as function of $h$ (to the right). Small (large) 
  values of $V$ ($h$) lead to large values of $R_{Jb}$.}
\label{fig:plot59}
\end{figure}

In Fig. (\ref{fig:plot59}) we show $R_{Jb}$ as function of 
$V = \sqrt{v_R^2+v_S^2}$ (left) and as function of $h$ (to the right).
The figure shows that large values of $R_{Jb}$ are obtained for 
small values of $V$ and for large values of $h$.  The decreasing 
of $R_{Jb}$ with increasing values of $V$ can be easily understood, 
since in the limit $V \to \infty$ the majoron should obviously 
decouple. 

\begin{figure}[htbp]
 \vspace{5mm}
   \begin{center}
   \includegraphics[width=75mm,height=50mm]{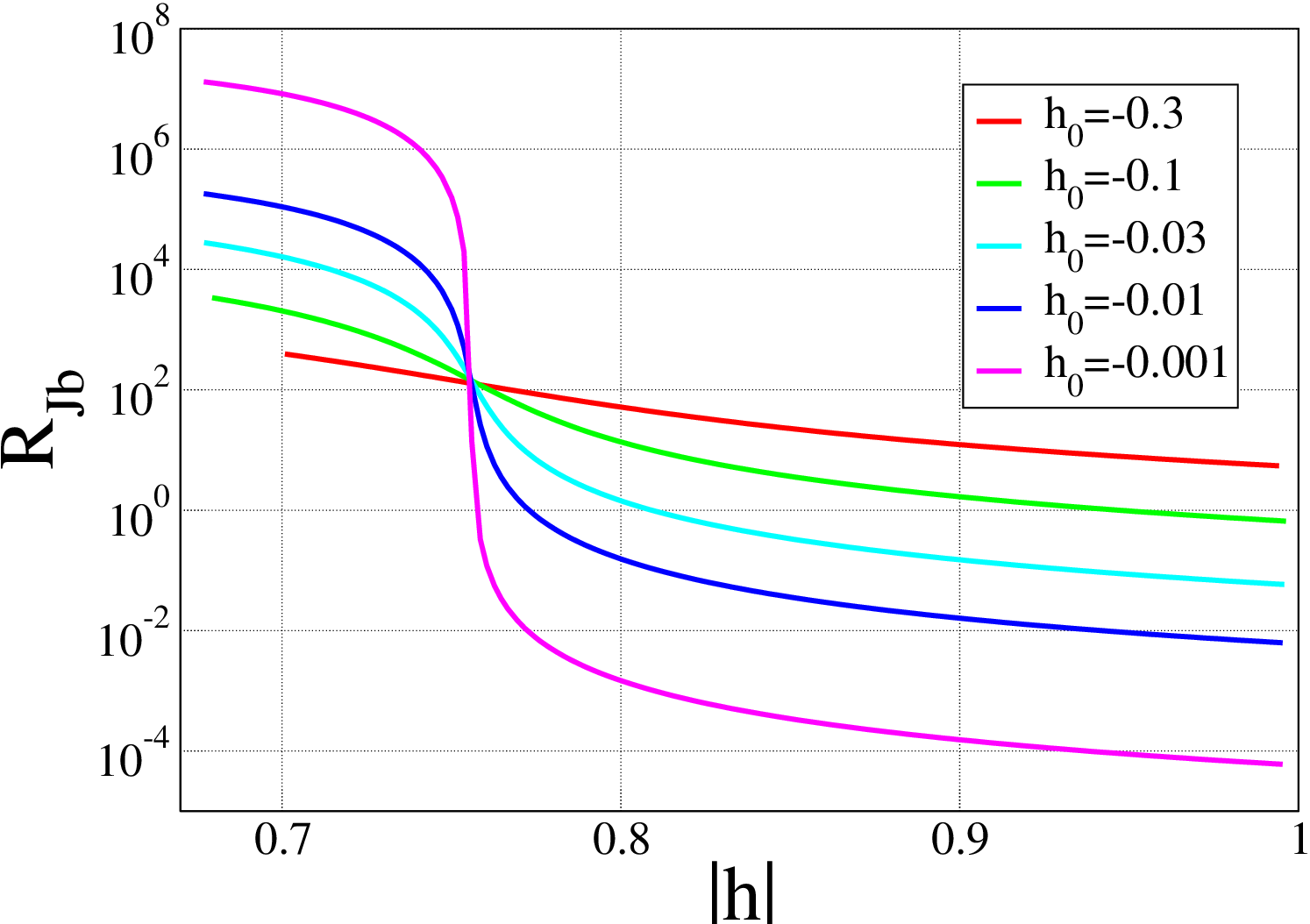}
\includegraphics[width=75mm,height=5cm]{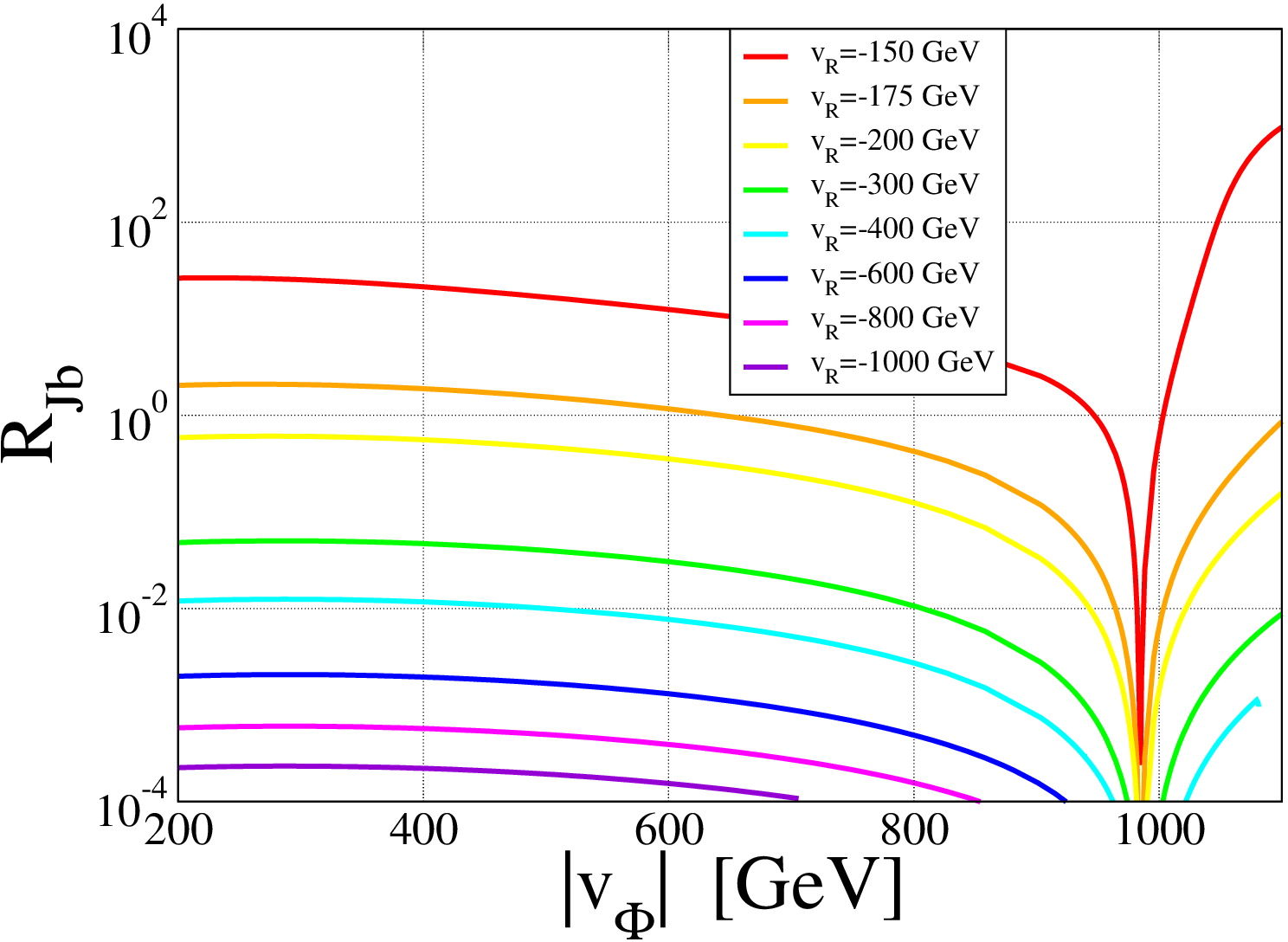}
   \end{center}
   \vspace{-5mm}
   \caption{Ratio $R_{Jb}$, defined in Eq.~(\ref{eq:ratio}), as 
     a) left figure: function of 
     $|h|$ for $-h_0=0.3,\, 0.1,\, 0.03,\, 0.01,\, 0.001$ 
     (on the right part of the plot from top to bottom). 
     The right panel b) gives $R_{Jb}$ as function of $|v_{\Phi}|$ for 
     different values of the parameter $v_R=v_S$ for $-v_R=
     150,\,175,\,200,\,300,\,400,\,600,\,800,\,1000$ GeV.}
   \label{fig:plot60}
   \end{figure}

Other singlet-sector parameters also can have an important impact on
$R_{Jb}$, as demonstrated in Fig. (\ref{fig:plot60}). As shown in the
left panel of this figure, larger values of $h_0$ lead to larger
values of $R_{Jb}$. For values of $h$ smaller than about $h \simeq
0.75$ (for our specific choice of the other parameters) the order of
the lines is exchanged. This is due to a level-crossing in the
eigenvalues. Below this value, the lightest Higgs is mainly a singlet
and thus even though it decays dominantly invisibly its production
cross section is very much reduced.

On the other hand, the right panel of Figure (\ref{fig:plot60}) 
shows that the value of $v_{\Phi}$ is normally somewhat less important
than the value of $V$ in determining $R_{Jb}$.  Again this can be
qualitatively understood since $V$ is the parameter whose magnitude
determines the breaking of lepton number (indeed, with the help of the
approximate couplings $g'_{i}$ in Eq.~(\ref{eq:16}) one can see that
the parameters $h$, $h_0$, $v_R$ and $v_S$ should be the most
important ones).

   As a summary of this section we conclude that large branching
   ratios of the doublet--like Higgs boson into invisible final states
   are possible in the SBRP model, despite the smallness of the
   neutrino masses indicated by oscillation data. Large values of
   $R_{Jb}$ occur for large values of the Yukawa couplings and
   for small values of $v_R$. The presence of the field $\Phi$ plays a
   crucial role in getting the invisible Higgs boson decays that are
   not suppressed by the small neutrino masses.

\subsection{Cubic--only superpotential}
\label{sec:cubic-only-superp}

Before concluding we illustrate the results we have obtained for the
case of a restricted SBRP model described by the superpotential in
Eq.~(\ref{eq:Wsuppot}) containing only cubic
terms~\cite{Romao:1997xf}. The restricted model provides a potential
``solution'' to the $\mu$ problem in the context of spontaneous
R--parity violation. We give results for the same parameter choices as
above, except that no mass parameters are now present in the basic
superpotential.

\begin{figure}[htbp]
\begin{center}
\vspace{5mm}
\includegraphics[clip,width=75mm,height=5cm]{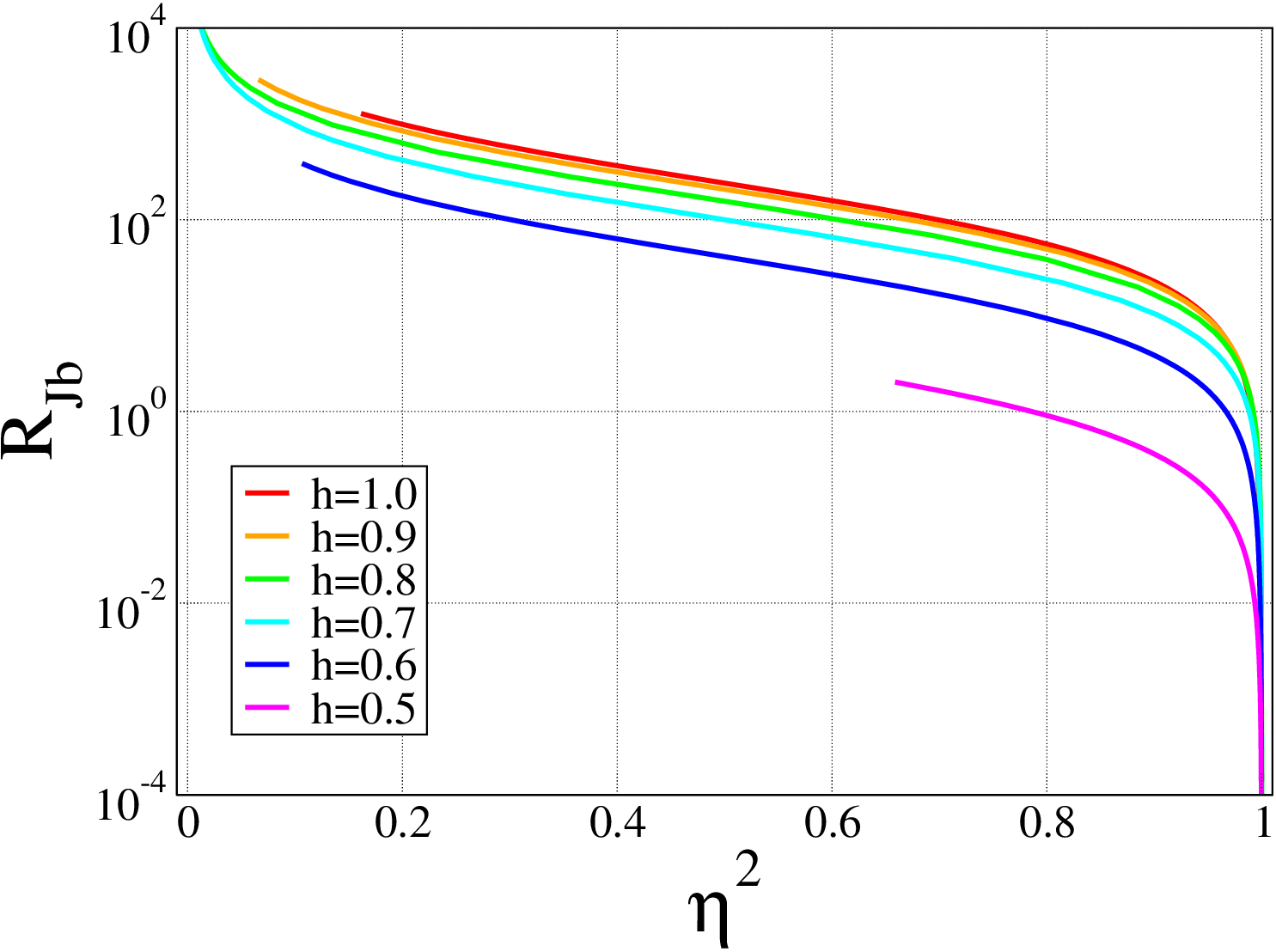}
\includegraphics[width=75mm,height=5cm]{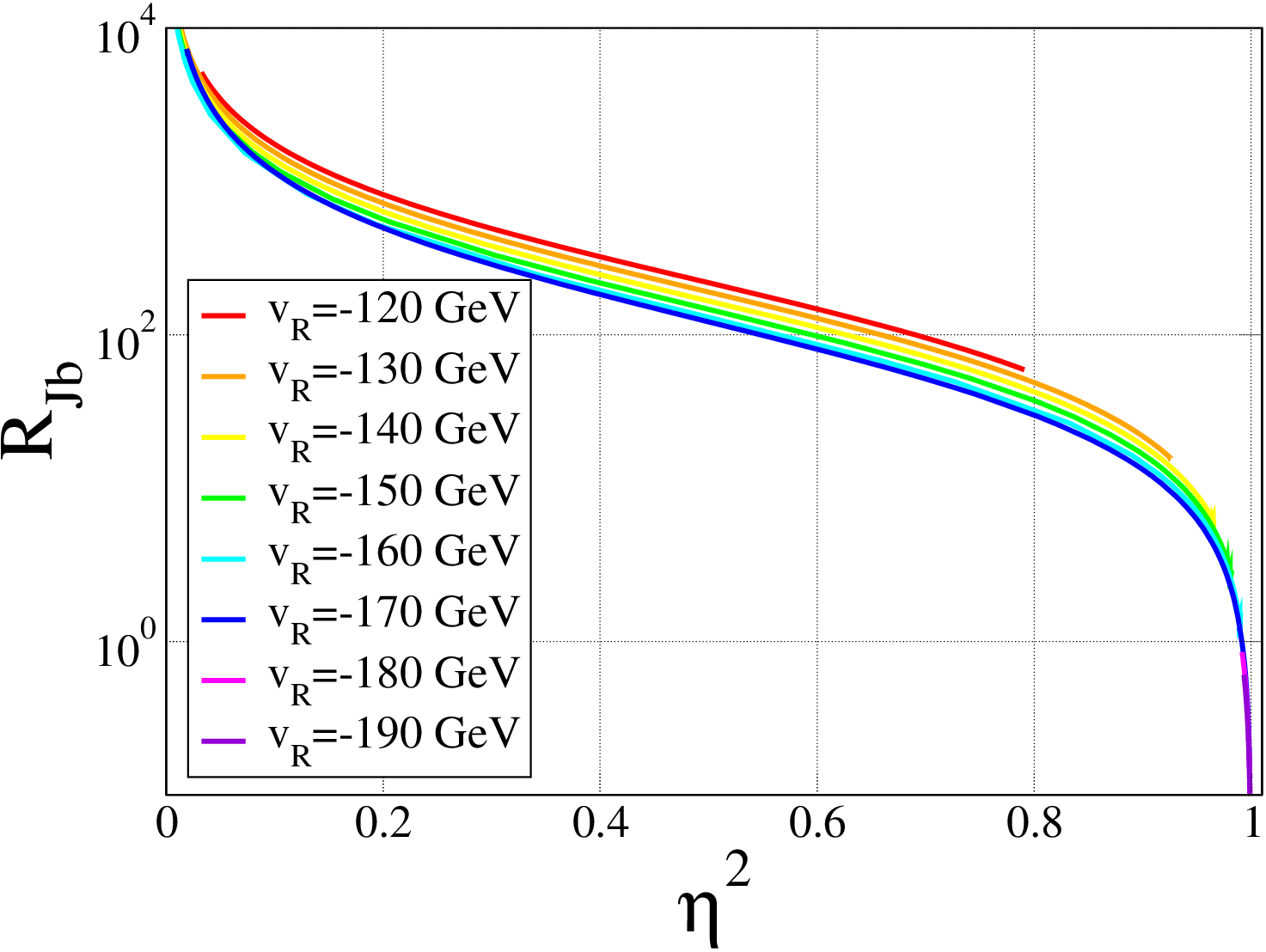}
\end{center}
\vspace{-5mm}
\caption{Ratio $R_{Jb}$, defined in Eq.~(\ref{eq:ratio}), as function 
  of $\eta^2$, a) to the left, for different values of the parameter
  $h$ and b) to the right, for different values of the parameter
  $v_R=v_S$. As in the general case (Fig.~\ref{fig:plot54}), large
  values of $R_{Jb}$ can be found even for $\eta\simeq 1$ also in the
  cubic-only case.}
\label{fig:plot143}
\end{figure}

\begin{figure}[htbp]
\begin{center}
\includegraphics[clip,width=75mm,height=5cm]{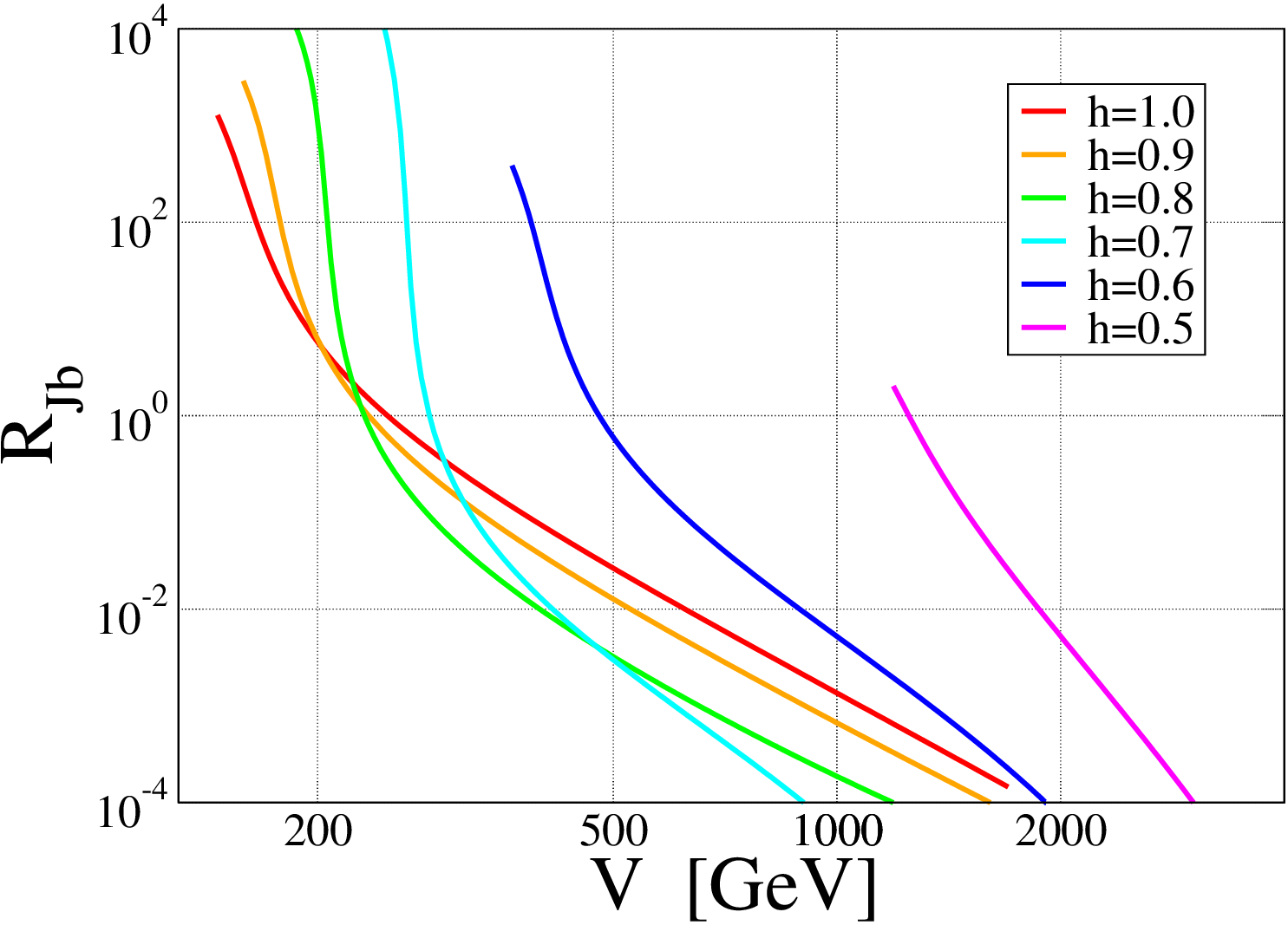}
\includegraphics[width=75mm,height=5cm]{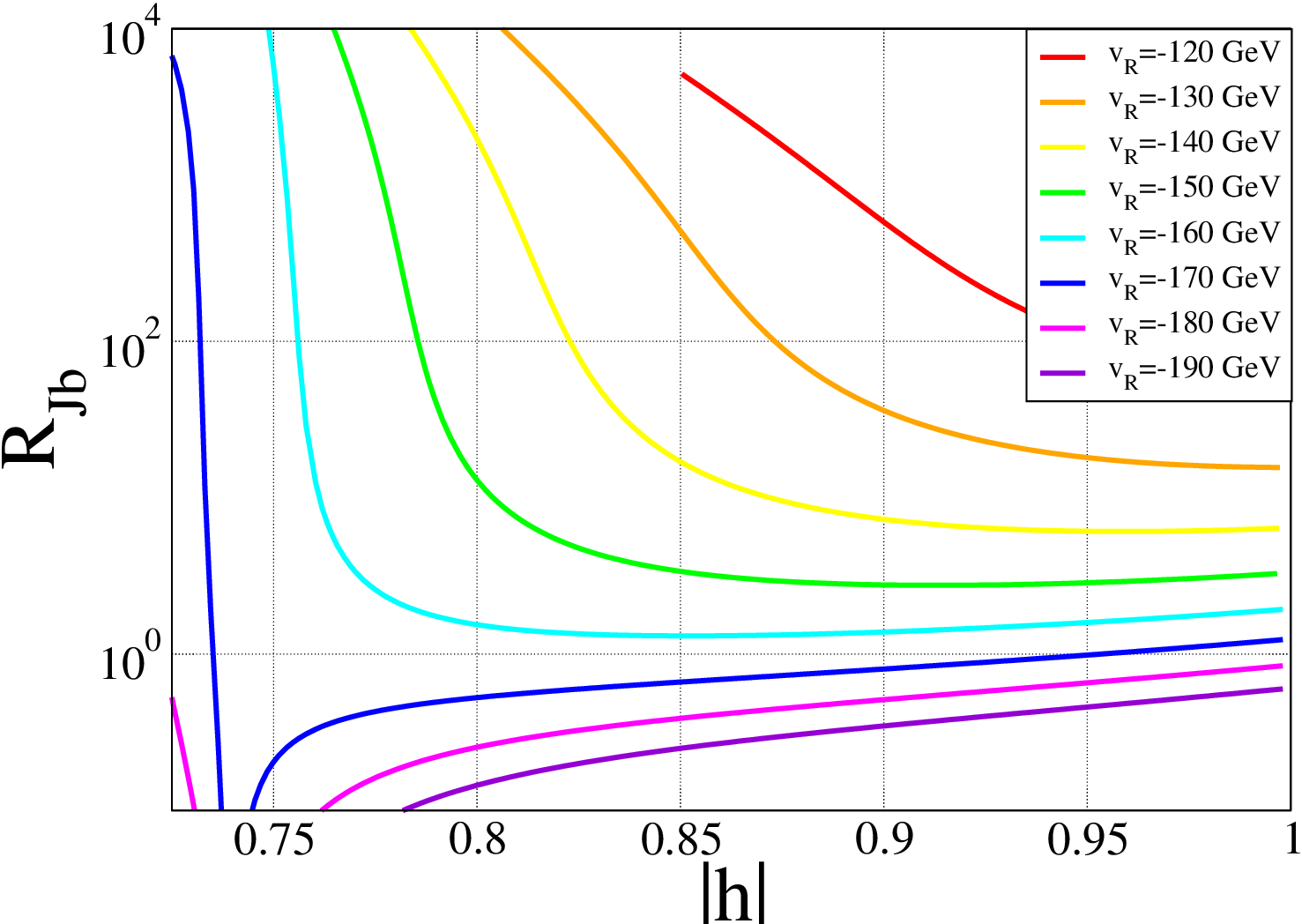}
\end{center}
\vspace{-5mm}
\caption{Ratio $R_{Jb}$, defined in Eq.~(\ref{eq:ratio}), as function 
  of the parameter $V$ (left) and as function of $h$ (right). The
  qualitative behaviour is similar to the general case, compare to
  figure \ref{fig:plot59}.}
\label{fig:plot144}
\end{figure}

Even though acceptable physical solutions consistent with experiment
(supersymmetric particle searches as well as neutrino oscillation
data) are somewhat harder to find, they exist. Figs.
(\ref{fig:plot143}) and (\ref{fig:plot144}) show $R_{Jb}$ as function
of $\eta^2$ and as function of $h$ and $V$ for the cubic-only case,
compare to Figs.  (\ref{fig:plot54}) and (\ref{fig:plot59}) for the
general case. As can be seen, the qualitative behaviour is very
similar in all cases, although the parameters for which acceptable
solutions are found are usually restricted to narrower ranges in the
cubic-only case. These figures demonstrate that also in the cubic-only
case large production cross section and large invisible branching
ratios for the lightest Higgs decay can occur at the same time.

\section{Discussion}
\label{sec:discussion}
      
We have discussed the possibility of an invisibly decaying Higgs boson in
the context of the spontaneously broken R--parity model.  One of the
neutral CP-odd scalars in this model corresponds to the \21 singlet
Nambu-Goldstone boson associated to the breaking of lepton number. In
contrast to the MSSM, where the Higgs boson can decay invisibly only
to supersymmetric states (in regions of parameters where the Higgs is
heavier than twice the lightest neutralino mass) in our case the Higgs
can decay mainly due to Eq.~(\ref{eq:HJJ}), instead of the Standard
Model channels, over large regions of parameters, given that there is
no kinematical barrier for this decay.  We have reanalysed this
striking suggestion in view of the recent data on neutrino
oscillations that indicate non-zero but small neutrino masses. We have
explicitly shown that (i) despite the smallness of neutrino masses,
invisible Higgs boson decay may indeed provide the most important mode
of Higgs boson decays and (ii) its production cross section need not
be suppressed with respect to that characterizing the standard MSSM
case. As a result, our analysis indicates that invisibly decaying
Higgs bosons should be an important topic in the agenda of future
accelerators, such as the Large Hadron Collider and the Next Linear
Collider. In fact the interest on this possibility goes beyond the
model we have taken as framework, it is much more general.  However
the SBRP model provides an attractive explanation for the origin of the 
neutrino masses that can also be probed at future collider experiments 
through the predicted pattern of the Lightest Supersymmetric Particle (LSP) 
decays which directly traces the experimentally observed neutrino mixing angles.

\section{Acknowledgments}

This work was supported by Spanish grant BFM2002-00345, by the
European Commission Human Potential Program RTN network
HPRN-CT-2000-00148 and by the European Science Foundation network
grant N.86.  M.H. is supported by a MCyT Ramon y Cajal contract.
A.~V.~M. was supported by a PhD fellowship from Generalitat
Valenciana. JCR was supported by the Portuguese 
\textit{Funda\c{c}\~ao para a Ci\^encia e a Tecnologia} 
under the contract CFIF-Plurianual and grant POCTI/FNU/4989/2002. 
We thank Werner Porod for useful discussions.

\appendix

\section{Neutrino-Neutralino-Singlino mass matrix}

In the basis 
\begin{equation}
(-i\lambda',-i\lambda^3,{\tilde H_d},{\tilde H_u},\nu_e,\nu_{\mu},\nu_{\tau},
\nu^c,S,\tilde{\Phi})
\label{eq:defbasis}
\end{equation}
the mass matrix of the neutral fermions following from Eq.~(\ref{eq:Wsuppot}) 
can be written as

\begin{equation}
\mathbf{M_N}=
\left(\begin{array}{lllll}
\mathbf{M_{\chi^0}} & \mathbf{m_{\chi^0\nu}}& \mathbf{m_{\chi^0\nu^c}}& 
\mathbf{0}& \mathbf{m_{\chi^0\Phi} } \\
\\
\mathbf{m^T_{\chi^0\nu}} & \mathbf{0} & \mathbf{m_{D}} & 
\mathbf{0} & \mathbf{0} \\
\\
\mathbf{m^T_{\chi^0\nu^c}}&\mathbf{m^T_{D}} & \mathbf{0} &
\mathbf{M_{\nu^c S}} & \mathbf{M_{\nu^c\Phi}} \\
\\
\mathbf{0} &\mathbf{0} &
\mathbf{M^T_{\nu^c S}} &\mathbf{0} &\mathbf{M_{S\Phi}} \\
\\
\mathbf{m^T_{\chi^0\Phi} } & \mathbf{0} & \mathbf{M^T_{\nu^c\Phi}} & 
\mathbf{M^T_{S\Phi}} & \mathbf{M_{\Phi}}
\end{array} \right).
\label{eq:mass}
\end{equation}
\noindent
where the matrix $\mathbf{M_{\chi^0}}$ is the MSSM neutralino mass
matrix:

\begin{equation}
\mathbf{M_{\chi^0}} =
\left(\begin{array}{llll}
M_1 & 0    & - \frac{1}{2} g' v_d & + \frac{1}{2} g' v_u \\ \vb{12}
0   & M_ 2 & + \frac{1}{2} g v_d &  - \frac{1}{2} g v_u \\ \vb{12}
- \frac{1}{2} g' v_d & + \frac{1}{2} g v_d &    0 & -\mu \\ \vb{12}
+ \frac{1}{2} g' v_u & - \frac{1}{2} g v_u & -\mu & 0
\end{array} \right).
\label{eq:mntrl}
\end{equation}
Here, $\mu ={\hat\mu}+h_0v_{\Phi}/\sqrt{2}$. 
$\mathbf{m_{\chi^0\nu}}$ is the R-parity violating neutrino-neutralino 
mixing part, which also appears in explicit bilinear R-parity breaking models:

\begin{equation}
\mathbf{m^T_{\chi^0\nu}} =
\left(\begin{array}{llll}
-\frac{1}{2}g'v_{L e} &\frac{1}{2}gv_{L e} & 0 & \epsilon_e \\[2mm]
-\frac{1}{2}g'v_{L \mu}& \frac{1}{2}gv_{L \mu}& 0& \epsilon_{\mu}\\[2mm]
-\frac{1}{2}g'v_{L \tau} & \frac{1}{2}gv_{L \tau} & 0& \epsilon_{\tau} 
\end{array} \right),
\label{eq:mrpv}
\end{equation}
where $v_{L i}$ are the vevs of the left-sneutrinos, $\epsilon_i$ 
are defined by $\epsilon_i = \frac{1}{\sqrt{2}}h_{\nu}^{i} v_R$, 
and $v_R$ is the vev of the right-sneutrino.

\noindent
Here $\mathbf{m_{\chi^0\nu^c}}$ is given as

\begin{equation}
\mathbf{m^T_{\chi^0\nu^c}} =
\left(
0, \hskip2mm 0, \hskip2mm  0, \hskip2mm 
\frac{1}{\sqrt{2}}\sum h_{\nu}^{i} v_{L i} 
\right).
\label{eq:mchinuc}
\end{equation}
and $\mathbf{m^T_{\chi^0\Phi} }$ is 

\begin{equation}
\mathbf{m^T_{\chi^0\Phi} }
= ( 0 , 0, - \frac{1}{\sqrt{2}}h_0 v_u , - \frac{1}{\sqrt{2}}h_0 v_d)
\label{eq:mchiphi}
\end{equation}
The ``Dirac'' mass matrix is defined in the usual way:

\begin{equation}
(\mathbf{m_{D}})_{i} = \frac{1}{\sqrt{2}}h_{\nu}^{i}v_u
\label{eq:mD}
\end{equation}
The $\nu^c$ and $S$ states are coupled by

\begin{equation}
(\mathbf{M_{\nu^c S}}) =  M_R + h\frac{v_{\Phi}}{\sqrt{2}}
\label{eq:mnucs}
\end{equation}
$\mathbf{M^T_{\nu^c\Phi}}$ and $\mathbf{M^T_{S\Phi}}$ are

\begin{equation}
\mathbf{M^T_{\nu^c\Phi}} = (\langle v_{S}\rangle)
\label{eq:mnucp}
\end{equation}
\begin{equation}
\mathbf{M^T_{S\Phi}} = (\langle v_{R}\rangle)
\label{eq:sp}
\end{equation}
Here, $\langle v_{R}\rangle = h v_R$ and $\langle v_{S}\rangle = h v_S$. 
Finally $\mathbf{M_{\Phi}}$ is 

\begin{equation}
\mathbf{M_{\Phi}} = M_{\Phi} + 
\lambda\frac{v_{\Phi}}{\sqrt{2}}
\label{eq:mp2}
\end{equation}

We briefly comment on the case of three generations of neutral
fermions in the singlet sector. For three copies of $\nu^c$ and $S$
fields the mass matrix of the neutral fermions can be written in
exactly the same form as given in Eq.~(\ref{eq:mass}) with some rather
straight-forward generalizations of the above definitions. These
changes are: $h$ and $h_{\nu}^i$ become $3 \times 3$ matrices $h^{ij}$
and $h_{\nu}^{ij}$. In Eq.~(\ref{eq:mchinuc}) the matrix becomes a $3
\times 4$ matrix, $M_R$ is a symmetric $3 \times 3$ matrix and Eqs.
(\ref{eq:mnucp}) and (\ref{eq:sp}) have to be replaced by

\begin{equation}
\mathbf{M^T_{\nu^c\Phi}} = (\langle v_{S_1}\rangle,\langle v_{S_2}\rangle ,
\langle v_{S_3}\rangle )
\end{equation}
\begin{equation}
\mathbf{M^T_{s\Phi}} = (\langle v_{R_1}\rangle,\langle v_{R_2}\rangle ,
\langle v_{R_3}\rangle )
\end{equation}
where $\langle v_{R i} \rangle = \sum_j h^{ji} v^R_j$ and 
$\langle v_{S_i} \rangle = \sum_j h^{ij} v^S_j$. 

Notice that even with three generations of $\nu^c$ and $S$ fields one
neutrino mass is zero at the tree-level.

\section{The Neutral Scalar Mass Matrix}

The $8\times 8$ scalar mass matrix is a symmetric matrix that in the
basis of the real part of 
$(H_d^0,H_u^0,\tilde\nu_i,\Phi,\tilde{S},\tilde\nu^c)$
can be written in the form,
\begin{eqnarray}
  \label{eq:1a}
  M^{S^2}=\left[
    \begin{array}{lll}
      M^{S^2}_{HH} & M^{S^2}_{H\widetilde L} & M^{S^2}_{HS}\\[+2mm]
      M^{S^2}_{H\widetilde L}\!{}^T & M^{S^2}_{\widetilde L \widetilde L} 
      & M^{S^2}_{\widetilde L S}\\[+2mm]
      M^{S^2}_{HS}\!{}^T &M^{S^2}_{\widetilde L S}\!{}^T & M^{S^2}_{SS}
    \end{array}
    \right]
\end{eqnarray}
where $M^{S^2}_{HH}$ is a symmetric $2\times 2$ matrix, $M^{S^2}_{\widetilde L
\widetilde L} $ and $M^{S^2}_{SS} $ are symmetric $3\times 3 $ matrices,
while   $M^{S^2}_{H\widetilde L}$ and $M^{S^2}_{HS}$ are $2\times 3$ matrices
and finally $M^{S^2}_{\widetilde L S}$ is (a non-symmetric) $3\times 3$
matrix. In this notation $\widetilde L$ denotes the sneutrinos and $S$
the singlet fields. 

We can write the mass matrix by giving the components of the various
blocks. We get,

\subsubsection*{$\bullet M^{S^2}_{HH}$}

\begin{eqnarray}
  \label{eq:2}
  M^{S^2}_{HH_{11}}&=&\frac{1}{4}(g^2+g'^2) v_d^2 + \Omega \tan\beta +
   \frac{\sqrt{2}}{2} \mu 
   \frac{v_R}{v_d} \sum_{i=1}^3 h_{\nu}^{i}\, v_{Li}\\[+2mm]
  M^{S^2}_{HH_{12}}&=&-\frac{1}{4}(g^2+g'^2) v_d v_u -\Omega +h_0^2 v_u v_d\\[+2mm]
  M^{S^2}_{HH_{22}}&=&\frac{1}{4}(g^2+g'^2) v_u^2 + \Omega \cot\beta -
   \frac{\sqrt{2}}{2} \frac{v_R}{v_u} \sum_{i=1}^3 
A_{h_{\nu}} h_{\nu}^{i}\, v_{Li}
-\frac{\sqrt{2}}{2}\, \widehat{M}_R \frac{v_S}{v_u} \sum_{i=1}^3 
   h_{\nu}^{i} v_{Li}
\end{eqnarray}
where,
\begin{eqnarray}
  \label{eq:3}
  \Omega&=& B \hat\mu
   -\delta^2 h_0 + \frac{\lambda}{4} h_0 v_{\Phi}^2+\frac{1}{2} h h_0
   v_R v_S + \frac{\sqrt{2}}{2} A_{h_0} h_0 v_{\Phi} +
  \frac{\sqrt{2}}{2} h_0 M_{\Phi} v_{\Phi}
\end{eqnarray}
and $\mu$, $\widehat{M}_R$ and $\widehat{M}_{\Phi}$ are defined 
in Eqs.~(\ref{eq:defmu}) and (\ref{eq:effmr}).

\subsubsection*{$\bullet M^{S^2}_{\widetilde L \widetilde L}$}

\begin{eqnarray}
  \label{eq:4}
  M^{S^2}_{\widetilde L \widetilde L_{ij}}&= &
\frac{1}{4}(g^2+g'^2) v_{Li} v_{Lj} + \frac{1}{2} \left(v_R^2+v_u^2\right) 
h_{\nu}^{i}
h_{\nu}^{j} + \delta_{ij} \left(
-\frac{\sqrt{2}}{2} \frac{v_u v_R}{v_{Li}}\, A_{h_{\nu}} h_{\nu}^{i}
\right.\nn\\ 
&\hskip-6mm& \left.
+ \frac{\sqrt{2}}{2} \frac{v_d v_R}{v_{Li}}\, h_{\nu}^{i}\, \mu 
- \frac{1}{2}\, \frac{v_R^2+v_u^2}{v_{Li}}\, h_{\nu}^{i}\, \sum_{k=1}^3
h_{\nu}^{k} v_{Lk} -\frac{\sqrt{2}}{2} \widehat{M}_R \frac{v_S v_u}{v_{Li}}
h_{\nu}^{i} \right)
\end{eqnarray}

\subsubsection*{$\bullet M^{S^2}_{\widetilde L S}$}

\begin{eqnarray}
  \label{eq:5}
  M^{S^2}_{\widetilde L S_{i1}}&=&
-\frac{1}{2}\, h_0 v_d\, v_R h_{\nu}^{i}
+\frac{1}{2}\, h\, v_u\, v_S h_{\nu}^{i} \\[+2mm]
M^{S^2}_{\widetilde L S_{i2}}&=& \frac{\sqrt{2}}{2} \widehat{M}_R\,
v_u  h_{\nu}^{i} \\[+2mm]
M^{S^2}_{\widetilde L S_{i3}}&=& 
\frac{\sqrt{2}}{2}\, v_u A_{h_{\nu}} h_{\nu}^{i} - 
\frac{\sqrt{2}}{2} h_{\nu}^{i} \mu
v_d +   h_{\nu}^{i} v_R  \sum_{k=1}^3
h_{\nu}^{k} v_{Lk}
\end{eqnarray}

\subsubsection*{$\bullet M^{S^2}_{H \widetilde L}$}

\begin{eqnarray}
  \label{eq:6}
  M^{S^2}_{H \widetilde L_{1i}}& = &
\frac{1}{4}(g^2+g'^2) v_d v_{Li} - \frac{\sqrt{2}}{2}\, \mu\, v_R\,
h_{\nu}^{i}\\[+2mm]
M^{S^2}_{H \widetilde L_{2i}}& = &
-\frac{1}{4}(g^2+g'^2) v_u v_{Li} + \frac{\sqrt{2}}{2}\, v_R\,
A_{h_{\nu}} h_{\nu}^{i} +\frac{\sqrt{2}}{2} \widehat{M}_R\, v_S h_{\nu}^{i} +
v_u\, h_{\nu}^{i}  \sum_{k=1}^3 h_{\nu}^{k} v_{Lk}
\end{eqnarray}

\subsubsection*{$\bullet M^{S^2}_{H S}$}

\begin{eqnarray}
  \label{eq:7}
  M^{S^2}_{H S_{11}}&=&\sqrt{2} h_0 \mu v_d  
  -\frac{\sqrt{2}}{2} h_0 \left(A_{h_0} + \widehat{M}_{\Phi}\right) v_u 
    - \frac{1}{2}
  h_0 v_R \sum_{k=1}^3  h_{\nu}^{k} v_{Lk}\\[+2mm]
  M^{S^2}_{H S_{12}}&=&-\frac{1}{2} h h_0\,
  v_R v_u \\[+2mm]
  M^{S^2}_{H S_{13}}&=&-\frac{1}{2} h h_0\,
  v_S v_u - \frac{\sqrt{2}}{2} \mu \sum_{k=1}^3  h_{\nu}^{k} v_{Lk}\\[+2mm]
  M^{S^2}_{H S_{21}}&=&\sqrt{2} h_0 \mu v_u  
  -\frac{\sqrt{2}}{2}h_0 \left(A_{h_0}+  \widehat{M}_{\Phi}\right) v_d
  +\frac{1}{2} h\, v_S \sum_{k=1}^3 h_{\nu}^{k} v_{Lk}
 \\[+2mm]
  M^{S^2}_{H S_{22}}&=&-\frac{1}{2} h h_0\, v_R v_d 
  +\frac{\sqrt{2}}{2} \widehat{M}_R \sum_{k=1}^3 h_{\nu}^{k} v_{Lk}
  \\[+2mm]
  M^{S^2}_{H S_{23}}&=&-\frac{1}{2} h h_0\,
  v_S v_d + v_u v_R \sum_{k=1}^3 h_{\nu}^{k} h_{\nu}^{k} +
\frac{\sqrt{2}}{2} \sum_{k=1}^3  A_{h_{\nu}} h_{\nu}^{k} v_{Lk}
\end{eqnarray}

\subsubsection*{$\bullet M^{S^2}_{S S}$}

\begin{eqnarray}
  \label{eq:8}
    M^{S^2}_{S S_{11}}&\hskip-3mm=\hskip -3mm& 
\frac{1}{2} \lambda^2 v_{\Phi}^2 + \delta^2  \left(
    C_{\delta} + M_{\Phi} \right) \frac{\sqrt{2}}{v_{\Phi}}
  -\frac{\sqrt{2}}{2} (v_d^2 + v_u^2) \frac{h_0 \hat\mu}{v_{\Phi}} +
\frac{\sqrt{2}}{4} \lambda \left( A_{\lambda} + 3  M_{\Phi}\right)
    v_{\Phi}\nn \\
&&  - \frac{\sqrt{2}}{2} h \left(A_h + M_{\Phi} \right)
    \frac{v_R v_S}{v_{\Phi}}
+ \frac{\sqrt{2}}{2} h_0 \left(A_{h_0} + M_{\Phi}\right)
\frac{v_u v_d }{v_{\Phi}} +
\frac{1}{2} h_0 \frac{v_d v_R}{v_{\Phi}} \sum_{k=1}^3 h_{\nu}^{k}
    v_{Lk}\nn \\
&& -\frac{1}{2} h\, \frac{v_S v_u}{v_{\Phi}}  \sum_{k=1}^3 h_{\nu}^{k} v_{Lk}
    -\frac{\sqrt{2}}{2} h\, M_R \frac{v_S^2 +v_R^2}{v_{\Phi}} 
\\[+2mm]
    M^{S^2}_{S S_{12}}&\hskip-3mm=\hskip -3mm& 
 \frac{\sqrt{2}}{2} h \left( A_h +\widehat{M}_{\Phi}\right) v_R 
 +\sqrt{2} h\, \widehat{M}_R v_S + \frac{1}{2} h\, v_u  \sum_{k=1}^3
 h_{\nu}^{k} v_{Lk} 
\\[+2mm]
    M^{S^2}_{S S_{13}}&\hskip-3mm=\hskip -3mm& 
 \frac{\sqrt{2}}{2} h \left( A_h + \widehat{M}_{\Phi}\right) v_S
-\frac{1}{2} h_0 v_d  \sum_{k=1}^3 h_{\nu}^{k} v_{Lk}
+\sqrt{2}\, h\, \widehat{M}_R v_R
\\[+2mm]
    M^{S^2}_{S S_{22}}&\hskip-3mm=\hskip -3mm& 
-\Gamma \frac{v_R}{v_S} - \frac{\sqrt{2}}{2} \frac{v_u}{v_S} 
\widehat{M}_R \sum_{k=1}^3 h_{\nu}^{k} v_{Lk}
\\[+2mm]
    M^{S^2}_{S S_{23}}&\hskip-3mm=\hskip -3mm& \Gamma + h^2 v_R v_S\\[+2mm]
    M^{S^2}_{S S_{33}}&\hskip-3mm=\hskip -3mm& -\Gamma \frac{v_S}{v_R} +
    \frac{\sqrt{2}}{2} \frac{\mu v_d}{v_R} \sum_{k=1}^3 h_{\nu}^{k}
    v_{Lk}  - \frac{\sqrt{2}}{2} \frac{v_u}{v_R} 
    \sum_{k=1}^3 A_{h_{\nu}} h_{\nu}^{k} v_{Lk}  
\end{eqnarray}
where
\begin{equation}
  \label{eq:9}
  \Gamma=B_{M_R} M_R -\delta^2 h + \frac{1}{4} h \lambda v_{\Phi}^2 -
  \frac{1}{2} h h_0 v_u v_d + \frac{\sqrt{2}}{2} h \left( A_h + 
  M_{\Phi}\right) v_{\Phi} 
\end{equation}

\section{The Neutral Pseudo--Scalar Mass Matrix}

The $8\times 8$ pseudoscalar mass matrix is a symmetric matrix that
can be written in the form,
\begin{eqnarray}
  \label{eq:11}
M^{P^2} = \left[
    \begin{array}{lll}
      M^{P^2}_{HH} & M^{P^2}_{H\widetilde L} & M^{P^2}_{HS}\\[+2mm]
      M^{P^2}_{H\widetilde L}\!{}^T & M^{P^2}_{\widetilde L \widetilde L} 
      & M^{P^2}_{\widetilde L S}\\[+2mm]
      M^{P^2}_{HS}\!{}^T &M^{P^2}_{\widetilde L S}\!{}^T & M^{P^2}_{SS}
    \end{array}
    \right]
\end{eqnarray}
where the blocks have the same structure as before.
We can write the mass matrix by giving the components of the various
blocks. We get,

\subsubsection*{$\bullet M^{P^2}_{HH}$}

\begin{eqnarray}
  \label{eq:12}
  M^{P^2}_{HH_{11}}&=&\Omega \tan\beta +
   \frac{\sqrt{2}}{2} \mu 
   \frac{v_R}{v_d} \sum_{i=1}^3 h_{\nu}^{i}\, v_{Li}\\[+2mm]
  M^{P^2}_{HH_{12}}&=&\Omega \\[+2mm]
  M^{P^2}_{HH_{22}}&=&\Omega \cot\beta -
   \frac{\sqrt{2}}{2} \frac{v_R}{v_u} \sum_{i=1}^3 A_{h_{\nu}} 
h_{\nu}^{i}\, v_{Li}
-\frac{\sqrt{2}}{2}\, \widehat{M}_R \frac{v_S}{v_u} \sum_{k=1}^3
   h_{\nu}^{k} v_{Lk} 
\end{eqnarray}
where $\Omega$ and $\mu$ are given in Eqs.~(\ref{eq:3}) and (\ref{eq:defmu}).

\subsubsection*{$\bullet M^{P^2}_{\widetilde L \widetilde L}$}

\begin{eqnarray}
  \label{eq:14}
\hskip -2mm
  M^{P^2}_{\widetilde L \widetilde L_{ij}}&=&
\frac{1}{2} \left(v_R^2+v_u^2\right) h_{\nu}^{i}
h_{\nu}^{j} + \delta_{ij} \left(
-\frac{\sqrt{2}}{2} \frac{v_u v_R}{v_{Li}}\, A_{h_{\nu}} h_{\nu}^{i} + 
\frac{\sqrt{2}}{2} \frac{v_d v_R}{v_{Li}}\, h_{\nu}^{i}\, \mu \right.\nn\\
&&\left.
- \frac{1}{2}\, \frac{v_R^2+v_u^2}{v_{Li}}\, h_{\nu}^{i}\, \sum_{k=1}^3
h_{\nu}^{k} v_{Lk} - \frac{\sqrt{2}}{2}\, \widehat{M}_R
  \frac{v_S v_u}{v_{Li}}h_{\nu}^{i} 
\right)
\end{eqnarray}

\subsubsection*{$\bullet M^{P^2}_{\widetilde L S}$}

\begin{eqnarray}
  \label{eq:15}
  M^{P^2}_{\widetilde L S_{i1}}&=&
-\frac{1}{2}\, h_0 v_d\, v_R h_{\nu}^{i} + \frac{1}{2}\, h\, v_u\,
v_S h_{\nu}^{i}\\[+2mm] 
M^{P^2}_{\widetilde L S_{i2}}&=& \frac{\sqrt{2}}{2} \widehat{M}_R\,
v_u\, h_{\nu}^{i}  \\[+2mm]
M^{P^2}_{\widetilde L S_{i3}}&=& 
-\frac{\sqrt{2}}{2}\, v_u A_{h_{\nu}} h_{\nu}^{i} + \frac{\sqrt{2}}{2} 
h_{\nu}^{i} \mu v_d 
\end{eqnarray}

\subsubsection*{$\bullet M^{P^2}_{H \widetilde L}$}

\begin{equation}
  \label{eq:16b}
  M^{P^2}_{H \widetilde L_{1i}}=- \frac{\sqrt{2}}{2}\, \mu\, v_R\,
h_{\nu}^{i},\hskip 10mm
M^{P^2}_{H \widetilde L_{2i}}=-\frac{\sqrt{2}}{2}\, v_R\,
A_{h_{\nu}} h_{\nu}^{i} - \frac{\sqrt{2}}{2}\, v_S \widehat{M}_R\, h_{\nu}^{i}
\end{equation}

\subsubsection*{$\bullet M^{P^2}_{H S}$}

\begin{eqnarray}
  \label{eq:17}
  M^{P^2}_{H S_{11}}&=&
\frac{\sqrt{2}}{2} h_0 \left(A_{h_0} -  \widehat{M}_{\Phi}\right) v_u 
    + \frac{1}{2}
  h_0 v_R \sum_{k=1}^3  h_{\nu}^{k} v_{Lk}\\[+2mm]
  M^{P^2}_{H S_{12}}&=&-\frac{1}{2} h h_0\,
  v_R v_u \\[+2mm]
  M^{P^2}_{H S_{13}}&=&-\frac{1}{2} h h_0\,
  v_S v_u - \frac{\sqrt{2}}{2} \mu \sum_{k=1}^3  h_{\nu}^{k} v_{Lk}\\[+2mm]
  M^{P^2}_{H S_{21}}&=&
  \frac{\sqrt{2}}{2}h_0 \left(A_{h_0}-  \widehat{M}_{\Phi}\right) v_d 
  + \frac{1}{2}\, h\, v_S \sum_{k=1}^3  h_{\nu}^{k} v_{Lk}
\\[+2mm]
  M^{P^2}_{H S_{22}}&=&-\frac{1}{2} h h_0\,
  v_R v_d +
  \frac{\sqrt{2}}{2}\widehat{M}_R  \sum_{k=1}^3  h_{\nu}^{k} v_{Lk} \\[+2mm]
  M^{P^2}_{H S_{23}}&=&-\frac{1}{2} h h_0\,
  v_S v_d  -
\frac{\sqrt{2}}{2} \sum_{k=1}^3  A_{h_{\nu}} h_{\nu}^{k} v_{Lk}
\end{eqnarray}

\subsubsection*{$\bullet M^{P^2}_{S S}$}

\begin{eqnarray}
  \label{eq:18}
    M^{P^2}_{S S_{11}}&\hskip-3mm=\hskip -3mm& 
\delta^2  \left(
    C_{\delta} + M_{\Phi} \right) \frac{\sqrt{2}}{v_{\Phi}}
  -\frac{\sqrt{2}}{2} (v_d^2 + v_u^2) \frac{h_0 \hat\mu}{v_{\Phi}} -
\frac{\sqrt{2}}{4} \lambda \left( 3 A_{\lambda} + M_{\Phi}\right)
    v_{\Phi} -2 B_{M_{\Phi}} M_{\Phi}\nn \\
&&  - \frac{\sqrt{2}}{2} h \left(A_h +  M_{\Phi} \right)
    \frac{v_R v_S}{v_{\Phi}}
+ \frac{\sqrt{2}}{2} h_0 \left(A_{h_0} + M_{\Phi}\right)
\frac{v_u v_d }{v_{\Phi}} +
\frac{1}{2} h_0 \frac{v_d v_R}{v_{\Phi}} \sum_{k=1}^3 h_{\nu}^{k}
    v_{Lk}\nn \\
&&+ 2 \delta^2 \lambda + \lambda h_0\, v_u v_d -
\lambda h\, v_R v_S -\frac{1}{2} h\, \frac{v_u v_S}{v_{\Phi}} 
\sum_{k=1}^3  h_{\nu}^{k} v_{Lk} - \frac{\sqrt{2}}{2} h\, M_R
\frac{v_S^2+v_R^2}{v_{\Phi}}  
\\[+2mm]
    M^{P^2}_{S S_{12}}&\hskip-3mm=\hskip -3mm& 
-\frac{\sqrt{2}}{2} h \left( A_h -  \widehat{M}_{\Phi}\right) v_R 
-\frac{1}{2} h\, v_u \sum_{k=1}^3  h_{\nu}^{k} v_{Lk}
\\[+2mm]
    M^{P^2}_{S S_{13}}&\hskip-3mm=\hskip -3mm& 
- \frac{\sqrt{2}}{2} h \left( A_h -  \widehat{M}_{\Phi}\right) v_S
-\frac{1}{2} h_0 v_d  \sum_{k=1}^3 h_{\nu}^{k}
    v_{Lk}\\[+2mm]
    M^{P^2}_{S S_{22}}&\hskip-3mm=\hskip -3mm& 
    -\Gamma \frac{v_R}{v_S}
    -\frac{\sqrt{2}}{2} \widehat{M}_R \frac{v_u}{v_S} 
    \sum_{k=1}^3  h_{\nu}^{k} v_{Lk}
    \\[+2mm]
    M^{P^2}_{S S_{23}}&\hskip-3mm=\hskip -3mm& -\Gamma \\[+2mm]
    M^{P^2}_{S S_{33}}&\hskip-3mm=\hskip -3mm& -\Gamma \frac{v_S}{v_R} +
    \frac{\sqrt{2}}{2} \frac{\mu v_d}{v_R} \sum_{k=1}^3 h_{\nu}^{k}
    v_{Lk}  - \frac{\sqrt{2}}{2} \frac{v_u}{v_R} 
    \sum_{k=1}^3 A_{h_{\nu}} h_{\nu}^{k} v_{Lk}  
\end{eqnarray}
where $\Gamma$ is given in Eq.~(\ref{eq:9}).


\begin{thebibliography}{10}

\bibitem{fukuda:1998mi}
Super-Kamiokande Collaboration, Y.~Fukuda {\em et~al.},
\newblock Phys. Rev. Lett. {\bf 81}, 1562 (1998), [hep-ex/9807003].

\bibitem{ahmad:2002jz}
SNO Collaboration, Q.~R. Ahmad {\em et~al.},
\newblock Phys. Rev. Lett. {\bf 89}, 011301 (2002), [nucl-ex/0204008].

\bibitem{eguchi:2002dm} T. Araki et al., [KamLAND Collaboration],
  arXiv:hep-ex/0406035. Talk by G. Gratta at Neutrino 2004, 14 19 June
  2004, Paris, http://neutrino2004.in2p3.fr/; K.~Eguchi {\em et~al.},
  \newblock Phys. Rev. Lett. {\bf 90}, 021802 (2003),
  [hep-ex/0212021].

\bibitem{Maltoni:2004ei} 
M.~Maltoni, T.~Schwetz, M.~A. Tortola and
  J.~W.~F. Valle, 
\newblock New J.\ Phys.\  {\bf 6}, 122 (2004), [hep-ph/0405172].
  See also M.~Maltoni et al, \newblock
  Phys. Rev. {\bf D68}, 113010 (2003), [hep-ph/0309130],
and references therein.

\bibitem{gell-mann:1980vs}
M.~Gell-Mann, P.~Ramond and R.~Slansky,
\newblock (1979),
\newblock Print-80-0576 (CERN).

\bibitem{yanagida:1979}
T.~Yanagida,
\newblock {\it In Proceedings of the Workshop on the Baryon Number of the Universe and Unified Theories, Tsukuba, Japan, 13-14 Feb 1979}.
\newblock {\it Edited by O.~Sawada and A.~Sugamoto}.

\bibitem{mohapatra:1981yp}
R.~N. Mohapatra and G.~Senjanovic,
\newblock Phys. Rev. {\bf D23}, 165 (1981).

\bibitem{chikashige:1981ui}
Y.~Chikashige, R.~N. Mohapatra and R.~D. Peccei,
\newblock Phys. Lett. {\bf B98}, 265 (1981).

\bibitem{schechter:1980gr}
J.~Schechter and J.~W.~F. Valle,
\newblock Phys. Rev. {\bf D22}, 2227 (1980).
\newblock Phys. Rev. {\bf D25}, 774 (1982).

\bibitem{mohapatra:1986bd}
R.~N. Mohapatra and J.~W.~F. Valle,
\newblock Phys. Rev. {\bf D34}, 1642 (1986).

\bibitem{zee:1980ai}
A.~Zee,
\newblock Phys. Lett. {\bf B93}, 389 (1980).

\bibitem{babu:1988ki}
K.~S. Babu,
\newblock Phys. Lett. {\bf B203}, 132 (1988).

\bibitem{valle:1991pk}
J.~W.~F. Valle,
\newblock Prog. Part. Nucl. Phys. {\bf 26}, 91 (1991).

\bibitem{Joshipura:1993hp}
A.~S. Joshipura and J.~W.~F. Valle,
\newblock Nucl. Phys. {\bf B397}, 105 (1993).

\bibitem{aulakh:1982yn}
C.~S.~Aulakh and R.~N.~Mohapatra,
\newblock Phys.\ Lett.\ B {\bf 119}, 136 (1982).
G.~G.~Ross and J.~W.~F.~Valle,
\newblock Phys.\ Lett.\ B {\bf 151}, 375 (1985).
J.~R.~Ellis {\it et al.},
\newblock Phys.\ Lett.\ B {\bf 150}, 142 (1985).
L.~J.~Hall and M.~Suzuki,
\newblock Nucl.\ Phys.\ B {\bf 231}, 419 (1984).

\bibitem{masiero:1990uj}
A.~Masiero and J.~W.~F. Valle,
\newblock Phys. Lett. {\bf B251}, 273 (1990).

\bibitem{romao:1992vu}
J.~C. Romao, C.~A. Santos and J.~W.~F. Valle,
\newblock Phys. Lett. {\bf B288}, 311 (1992).

\bibitem{shiraishi:1993di}
M.~Shiraishi, I.~Umemura and K.~Yamamoto,
\newblock Phys. Lett. {\bf B313}, 89 (1993).

\bibitem{diaz:1998xc}
M.~A. Diaz, J.~C. Romao and J.~W.~F. Valle,
\newblock Nucl. Phys. {\bf B524}, 23 (1998), [hep-ph/9706315].

\bibitem{chun:1999bq}
E.~J. Chun and S.~K. Kang,
\newblock Phys. Rev. {\bf D61}, 075012 (2000), [hep-ph/9909429].

\bibitem{abada:2001zh}
A.~Abada, S.~Davidson and M.~Losada,
\newblock Phys. Rev. {\bf D65}, 075010 (2002), [hep-ph/0111332].

\bibitem{Diaz:2003as}
M.~A. Diaz {\em et~al.},
\newblock Phys. Rev. {\bf D68}, 013009 (2003), [hep-ph/0302021].

\bibitem{hirsch:2000ef}
M.~Hirsch {\em et~al.},
\newblock Phys. Rev. {\bf D62}, 113008 (2000), [hep-ph/0004115],
\newblock Err-ibid. {\bf D65}:119901,2002.

\bibitem{romao:1999up}
J.~C. Romao {\em et~al.},
\newblock Phys. Rev. {\bf D61}, 071703 (2000), [hep-ph/9907499].

\bibitem{Hirsch:2004he}
M.~Hirsch and J.~W.~F. Valle, New J. Phys. 6 76 (2004),
\newblock hep-ph/0405015.

\bibitem{hirsch:2002ys}
M.~Hirsch {\em et~al.},
\newblock Phys. Rev. {\bf D66}, 095006 (2002), [hep-ph/0207334].

\bibitem{porod:2000hv}
W.~Porod {\em et~al.},
\newblock Phys. Rev. {\bf D63}, 115004 (2001), [hep-ph/0011248].

\bibitem{restrepo:2001me}
D.~Restrepo, W.~Porod and J.~W.~F. Valle,
\newblock Phys. Rev. {\bf D64}, 055011 (2001), [hep-ph/0104040].

\bibitem{Hirsch:2003fe}
M.~Hirsch and W.~Porod,
\newblock Phys. Rev. {\bf D68}, 115007 (2003), [hep-ph/0307364].

\bibitem{deCampos:1995av}
F.~de~Campos {\em et~al.},
\newblock Nucl. Phys. {\bf B451}, 3 (1995), [hep-ph/9502237].

\bibitem{Panagiotakopoulos:1998yw}
C.~Panagiotakopoulos and K.~Tamvakis,
\newblock Phys. Lett. {\bf B446}, 224 (1999), [hep-ph/9809475].

\bibitem{romao:1992zx}
J.~C. Romao, F.~de~Campos and J.~W.~F. Valle,
\newblock Phys. Lett. {\bf B292}, 329 (1992), [hep-ph/9207269].

\bibitem{Barbieri:1982eh}
R.~Barbieri, S.~Ferrara and C.~A. Savoy,
\newblock Phys. Lett. {\bf B119}, 343 (1982).

\bibitem{abreu:2000mm}
DELPHI Collaboration, P.~Abreu {\em et~al.},
\newblock Phys. Lett. {\bf B502}, 24 (2001), [hep-ex/0102045].

\bibitem{Romao:1997xf}
J.~C.~Romao, A.~Ioannisian and J.~W.~F.~Valle,
\newblock Phys.\ Rev.\ D {\bf 55} (1997) 427
[arXiv:hep-ph/9607401].

\bibitem{lopez-fernandez:1993tk}
A.~Lopez-Fernandez {\em et~al.}, 
\newblock Phys. Lett. {\bf B312}, 240 (1993), [hep-ph/9304255].

\bibitem{deCampos:1997bg}
F.~de~Campos {\em et~al.},
\newblock Phys. Rev. {\bf D55}, 1316 (1997), [hep-ph/9601269].

\bibitem{Abdallah:2003ry}
J.~Abdallah {\it et al.}  [DELPHI Collaboration],
Eur.\ Phys.\ J.\ C {\bf 32} (2004) 475
[arXiv:hep-ex/0401022].

\bibitem{Allanach:2002nj}
B.~C. Allanach {\em et~al.},
\newblock Eur. Phys. J. {\bf C25}, 113 (2002), [hep-ph/0202233].

\bibitem{SPheno}W. Porod, Comput. Phys. Commun.  153 (2003) 275 
[arXiv:hep-ph/0301101


\end{thebibliography}

\end{document}